\documentclass{article}

\usepackage{arxiv}

\usepackage[utf8]{inputenc} 
\usepackage[T1]{fontenc}    
\usepackage{hyperref}       
\usepackage{url}            
\usepackage{booktabs}       
\usepackage{amsfonts}       
\usepackage{nicefrac}       
\usepackage{microtype}      
\usepackage{lipsum}		
\usepackage{graphicx}
\usepackage{natbib}
\usepackage{doi}
\usepackage{amsmath}
\usepackage{bbm}
\usepackage[usenames,dvipsnames,svgnames,table]{xcolor}
\usepackage{hyperref}
\hypersetup{
     colorlinks   = true,
     citecolor    = teal
}

\title{Joint Modeling of Multiple Longitudinal Biomarkers and Survival Outcomes via Threshold Regression: Variability as a Predictor}

\author{ \href{https://orcid.org/0009-0009-6361-9760}{\includegraphics[scale=0.06]{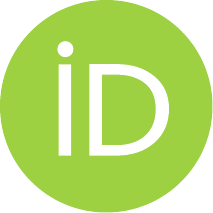}\hspace{1mm}Mingyan~Yu} \\
	Department of Biostatistics\\
	University of Michigan\\
	Ann Arbor, MI 48109 \\
	\texttt{myanyu@umich.edu} \\
	\And
	\href{https://orcid.org/0000-0001-7582-669X}{\includegraphics[scale=0.06]{orcid.pdf}\hspace{1mm}Zhenke~Wu} \\
	Department of Biostatistics\\
	University of Michigan\\
	Ann Arbor, MI 48109 \\
	\texttt{zhenkewu@umich.edu} \\
    \And Michelle M.~Hood \\
	Department of Epidemiology\\
	University of Michigan\\
	Ann Arbor, MI 48109 \\
	\texttt{mmhood@umich.edu} \\
    \And Carrie A. Karvonen-Gutierrez \\
	Department of Epidemiology\\
	University of Michigan\\
	Ann Arbor, MI 48109 \\
	\texttt{ckarvone@umich.edu} \\
    \And Siob\'an D.~Harlow \\
	Department of Epidemiology\\
	University of Michigan\\
	Ann Arbor, MI 48109 \\
	\texttt{harlow@umich.edu} \\
    \And 
    \href{https://orcid.org/0000-0003-4128-5527}{\includegraphics[scale=0.06]{orcid.pdf}\hspace{1mm}Michael R.~Elliott} \\
	Department of Biostatistics\\
	University of Michigan\\
	Ann Arbor, MI 48109 \\
	\texttt{mrelliot@umich.edu} \\
}

\renewcommand{\shorttitle}{Joint Modeling of Multiple Longitudinal Biomarkers and Survival Outcomes}

\hypersetup{
pdftitle={Joint Modeling of Multiple Longitudinal Biomarkers and Survival Outcomes},
pdfsubject={Joint Model},
pdfauthor={Mingyan~Yu, Zhenke~Wu, Michelle M.~Hood, Carrie Karvonen-Gutierrez, Siob\'an D. Harlow, Michael R.~Elliott},
pdfkeywords={Bayesian hierarchical model, Biomarker variability, Joint multivariate modeling, Limit of detection, Women's Health},
}

\begin{document}
\maketitle

\begin{abstract}
	Longitudinal biomarker data and health outcomes are routinely collected in many studies to assess how biomarker trajectories predict health outcomes. Existing methods primarily focus on mean biomarker profiles, treating variability as a nuisance. However, excess variability may indicate system dysregulations that may be associated with poor outcomes. In this paper, we address the long-standing problem of using variability information of multiple longitudinal biomarkers in time-to-event analyses by formulating and studying a Bayesian joint model. We first model multiple longitudinal biomarkers, some of which are subject to limit-of-detection censoring. We then model the survival times by incorporating random effects and variances from the longitudinal component as predictors through threshold regression that admits non-proportional hazards. We demonstrate the operating characteristics of the proposed joint model through simulations and apply it to data from the Study of Women’s Health Across the Nation (SWAN) to investigate the impact of the mean and variability of follicle-stimulating hormone (FSH) and anti-M\"ullerian hormone (AMH) on age at the final menstrual period (FMP).
\end{abstract}


\section{Introduction}
\label{sec:intro}
The final menstrual period (FMP), defined as the last menstrual period preceding 12 consecutive months of amenorrhea without other pathological or physiological causes \citep{who1996research}, serves as a biological signal of ovarian aging and natural menopause. Early menopause is associated with an increased risk of mortality from cardiovascular disease and an earlier decline in cognitive function \citep{gold2011timing}, making the prediction of the age at FMP crucial for timely interventions. Prior research has identified changes in several hormones during the menopausal transition (MT), e.g., follicle-stimulating hormone (FSH) increases \citep{freeman2005follicular,randolph2011change} and anti-M\"ullerian hormone (AMH) declines and becomes undetectable by late MT and postmenopause \citep{van2004anti}. However, the joint effect of FSH and AMH on age at FMP has not been estimated, nor has the potential impact of \underline{variability} in these hormones been considered. Toward this goal, a statistical method that integrates both the mean trends and variability of multiple biomarkers with time-to-event outcomes is required. 

Because of the inherent measurement error in longitudinal biomarker data \citep{wang2020methods}, we employ joint modeling of longitudinal and survival data. Previous work on this topic has commonly modeled the longitudinal data by mixed-effects models \citep{papageorgiou2019overview} or latent class models \citep{proust2014joint} with random effects or latent classes incorporated as predictors in a Cox proportional hazards model for survival outcomes. However, the classical Cox proportional hazards model can sometimes be problematic since the proportional hazards assumption may not always hold for various reasons. Violation of such assumption may lead to biased effect estimates and decreased statistical power \citep{schemper1992cox}.
Some common remedies include incorporating time-dependent covariate effects, covariate stratification, dividing time into intervals that fulfill the assumption and using weighted estimation for Cox regression \citep{schemper2009estimation}. Here we undertake a novel application of threshold regression \citep{lee2006threshold}. Threshold regression models survival times using latent stochastic processes, in which an event occurs when the latent process crosses a threshold for the first time (``first-hitting time''), shown to relax the proportional hazards assumption. 

The second major contribution is considering the variability of multiple biomarkers as a predictor in the time-to-event model. While much attention has been given to the impact of the mean biomarker values in prior research, variability is often treated as a nuisance parameter. However, within-subject biomarker variability can be of vital importance under some circumstances. For example, \citet{gao2011joint} demonstrated that the intraocular pressure variability is independently predictive of primary open-angle glaucoma through a joint modeling framework. Similarly, \citet{wang2024modeling} found that higher systolic blood pressure variability significantly increases cardiovascular event risk. Despite these findings, most of such studies involve only a single biomarker. Although some efforts have been made to include multiple biomarkers \citep{chi2006joint, rizopoulos2011bayesian, long2018joint}, those studies often overlook the predictability of the biomarker variability. 

In addition to extending the joint model to incorporate multiple biomarkers and using their variabilities as predictors, we also need to deal with biomarkers with detection limits, which is a challenge in our motivating application. The assay used to measure AMH has a limit of detection (LOD) and AMH becomes undetectable and left-censored when it falls below a certain threshold. Simple conventional methods include complete-case analysis or imputing left-censored biomarkers with constant values such as LOD/2 or LOD/$\sqrt{2}$ \citep{schisterman2006limitations}, which typically introduce bias in subsequent analysis, especially when the proportion of censored data is moderate to large \citep{liu2013two}. We choose to adopt the Tobit regression-based imputation method introduced by \citet{liu2013two} within our joint model framework to generate more reliable imputed values and parameter estimates. 

The remainder of our manuscript is organized as follows: Section \ref{sec:threg} introduces the basic concepts of the first-hitting-time approach and threshold regression; Section \ref{sec:method} develops the proposed joint model in a Bayesian framework; Section \ref{sec:simulation} conducts simulation studies to assess the operating characteristics of the joint model and compare the performance against common alternatives; Section \ref{sec:application} presents analysis results using the SWAN data; Section \ref{sec:discussion} concludes with a summary and discussion of potential limitations and future directions.

\section{First-Hitting Time Model and Threshold Regression}
\label{sec:threg}

The first-hitting-time (FHT) model consists of two key components. The first component, represented by the stochastic process \{$Y(t)$\}, models a subject's disease progression or health status. This stochastic process is initiated at $y_0$ when $t=0$ and evolves over time. Let $\mathcal{T}$ represent the time space of the stochastic process and $\mathcal{Y}$ the state space of the process. The second component of the FHT model, represented by a set $\mathcal{B}$, defines the boundary set or threshold and is a subset of $\mathcal{Y}$. The boundary set $\mathcal{B}$ can be either fixed or time-dependent. Here we assume a fixed boundary for all subjects. Assuming the initial status $y_0$ does not lie within $\mathcal{B}$, then the parameter of interest is the first time the stochastic process \{$Y(t)$\} enters or hits the boundary $\mathcal{B}$, denoted by $S=\text{inf} \{t: Y(t)\in \mathcal{B}\}$. Under most circumstances, we can only observe the first hitting time $S$, while the underlying process $\{Y(t)\}$ remains latent and unobservable. 

Several stochastic processes can be applied in FHT modeling, with one of the most widely used being the Wiener process. The Wiener process is a random walk in continuous time and space, more specifically, a Brownian motion with drift \citep{caroni2017first}. The Wiener process allows for both improvement and deterioration of the latent stochastic process over time due to its normally distributed increments, making it especially suitable for modeling human health trajectories, where fluctuations are common. The Wiener process has been applied to numerous studies. One of the earliest applications is by \citet{lancaster1972stochastic}, who studied the duration of a strike through an FHT model. The method has been extended in several ways over the last two decades. \citet{pennell2010bayesian} incorporated random effects to account for between-subject heterogeneity. \citet{economou2015bayesian} extended the model to accommodate recurrent events. The model has also been adapted to the context of high-dimensional data \citep{de2023boosting}. Another frequently used continuous-time process is the Gamma process. However, unlike the Wiener process, the FHT model with Gamma process is monotonic and no improvement is allowed, making it more suitable for modeling degradation in materials and quality of systems \citep{lawless2004covariates, park2005accelerated, ling2014accelerated}. The Poisson and Bernoulli processes are similar in that the observed event is triggered by a specific number of occurrences; however, the former models continuous time while the latter models discrete time. In this paper, given that our case study targets a health-related research question, we employ the Wiener process for its capacity to capture fluctuations inherent in health status. 

Let $\{Y_i(t)\}$ denote the latent health status of subject $i$ at time $t$ under a Wiener process characterized by parameters $y_{0i}$, $\zeta_i$ and $\sigma^2$. Here, $y_{0i}$ denotes the initial health status of subject $i$ at $t=0$, $\zeta_i$ is the drift parameter quantifying the rate at which subject $i$ approaches the threshold triggering the event, and $\sigma^2$ is the variance of the latent stochastic process. Let $\{W(t)\}$ denote the standard Wiener process with the following properties:
\begin{itemize}
    \item $W(0)=0$ almost surely;
    \item Independent increments: for every $t_{j}>0$, the future increments $W(t_{j+u})-W(t_j)$, $u\geq0$, are independent of the past values $W(t_k)$, $k<j$;
    \item Gaussian increments: for every $t_j>0$, $W(t_{j+u})-W(t_j)\sim \mathcal{N}(0, t_{j+u}-t_{j}), u\geq0$;
\end{itemize}
The stochastic process is then given by $Y_i(t)=y_{0i}+\zeta_it+\sigma W(t)$. In our case, the boundary set $\mathcal{B}$ is defined as $(-\infty,0]$ and an event occurs when $Y_i(t)\le 0$ for the first time, assuming $y_{0i}>0$. Under such assumption, the first-hitting-time $S_i$ for individual $i$ follows an inverse-Gaussian distribution with mean $-y_{0i}/\zeta_i$ and shape $y_{0i}^2/\sigma^2$ and probability density function
\begin{equation}
    \begin{gathered}
    f(S_i|y_{0i},\zeta_i,\sigma^2)=\frac{y_{0i}}{\sqrt{2\pi\sigma^2S_i^3}}\exp\left(-\frac{(y_{0i}+\zeta_iS_i)^2}{2\sigma^2S_i}\right).
\end{gathered}\label{eq:3.1}
\end{equation}
Since only two free parameters are involved in the inverse-Gaussian distribution, a common practice is to set the variance parameter $\sigma^2$ to 1 \citep{lee2008threshold}. Then both the initial status $y_{0i}$ and drift $\zeta_i$ will be measured in units of the standard deviation of the process. In threshold regression, covariate information is further incorporated and associated with the two latent components $y_{0i}$ and $\zeta_i$ through regression models. 

\begin{figure}
\begin{center}
\includegraphics[width=3in]{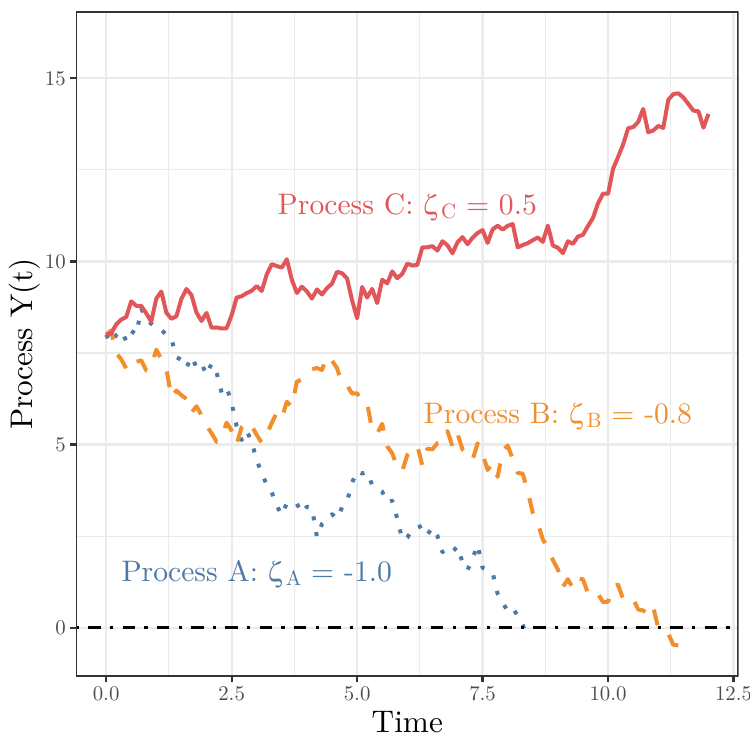}
\end{center}
\caption{\small Demonstration of three sample stochastic processes with the same initial status and different drift values and signs. An event occurs when the process reaches the threshold at 0, represented by the dot-dashed horizontal line.
Process A, represented by the dotted line with the most negative drift, experiences the earliest event occurrence, whereas Process C, a potential ``cured" case represented by the solid line with a positive drift, moves away from the threshold over time. 
 \label{fig:TR_demo}}
\end{figure}

Theoretically, the latent process $\{Y_i(t)\}$ will eventually reach 0 and the event occurs when $\zeta_i$ is non-positive. However, we do not impose any constraints on $\zeta_i$ and non-positivity is therefore not guaranteed. When $\zeta_i>0$, there is a probability that the event will never occur, and we refer to this probability as the ``cure rate" with $\mathcal{P}(\text{cure}_i)=\mathcal{P}(S_i=\infty)=1-\exp\left({2y_{0i}\zeta_i/\sigma^2}\right)$. Figure \ref{fig:TR_demo} illustrates three sample latent processes with the same initial status but different drift parameter values and signs, demonstrating how an event occurs and how the latent process behaves for a potential ``cured" case. However, since we are modeling age at FMP in our application, it is unlikely for us to observe a ``cured" subject as menopause is a physiological change that every female will experience in our motivating study.

\section{Methods}
\label{sec:method}

\subsection{Notation}
\label{sec3.1}

Let the censoring indicator $\delta_i=1$ if the event time $T_i$ is observed and $\delta_i=0$ for right-censored event where $C_i$ is observed instead. Let matrix $\boldsymbol{X}_i$ of dimension $n_i \times Q$ represent the observed biomarker values of subject $i$ over $n_i$ visits at time $\boldsymbol{t}_i=(t_{i1},\ldots,t_{in_i})^\top$. Each row of $\boldsymbol{X}_i$ corresponds to a visit, with the $j$-th row, $\boldsymbol{X}_{ij}=(X_{ij1},
\ldots,X_{ijQ})$ representing the values of $Q$ biomarkers measured at time $t_{ij}$. The vector $\boldsymbol{Z}_i$ captures baseline covariates and $\boldsymbol{L}$ is a $l$-dimensional vector of detection limits for $l\le Q$ biomarkers that are subject to a limit of detection; here we assume a common limit of detection across all subjects. Let ${\mathcal{D}}_i=(T_i, C_i, \delta_i, \boldsymbol{X}_i, \boldsymbol{t}_i, \boldsymbol{Z}_i, \boldsymbol{L})$ denote the observed data for subject $i$, $i=1,\ldots,N$. 

\subsection{Proposed Joint Modeling Framework}
\label{sec3.2}

Our proposed joint modeling approach is comprised of two connected components. We first specify regression models with mixed effects for longitudinal biomarkers ($t_{ij}, \boldsymbol{X}_{ij}$) and perform imputation for biomarkers below detection limits; then we link the time-to-event outcome $T_i$ with the individual-specific random effects and variances from the longitudinal model along with baseline covariates $\boldsymbol{Z}_i$ to draw inferences on how the mean and variability of biomarkers affect the time-to-event outcomes.

\subsubsection{Component 1: Multivariate Longitudinal Biomarkers}
\label{sec3.2.1}

We specify the mixed-effect regression model for the longitudinal biomarkers as follows:
\begin{equation}
\begin{gathered}
\boldsymbol{X}_{ij}|\boldsymbol{\beta}_1,\ldots, \boldsymbol{\beta}_Q, \boldsymbol{B}_i, \boldsymbol{S}_i \sim \mathcal{N}_Q(\boldsymbol{\mu}_{ij}=\boldsymbol{\mu}(t_{ij};\boldsymbol{\beta}_1,\ldots, \boldsymbol{\beta}_Q, \boldsymbol{B}_i), \boldsymbol{S}_i), \ j=1,\ldots,n_i; \\
    \boldsymbol{b}_{iq}\sim \mathcal{N}_P (\boldsymbol{0}, \boldsymbol{\Sigma}_q), \ q=1,\ldots,Q, \ \text{ independently for } i=1,\ldots,N,
\end{gathered}\label{eq:3.2}
\end{equation}
where $\mathcal{N}_Q(\boldsymbol{\mu}_{ij}, \boldsymbol{S}_i)$ represents a $Q$-dimensional multivariate Gaussian distribution with mean vector $\boldsymbol{\mu}_{ij}$ and variance-covariance matrix $\boldsymbol{S}_i$. $\boldsymbol{B}_i=(\boldsymbol{b}_{i1},\ldots,\boldsymbol{b}_{iQ})$ is a $P\times Q$ matrix of random effects, and for simplicity, we assume the same number of $P$ random effects across all biomarkers. The $q$-th column of $\boldsymbol{B}_i$, $\boldsymbol{b}_{iq}=(b_{iq1},\ldots,b_{iqP})^\top$, represents the random effects for the $q$-th biomarker where $P$ is the number of basis functions of time; in our motivating application, we will only include random intercepts and slopes for interpretation purpose and due to the limited number of measurements we have on each biomarker. The mean function $\boldsymbol{\mu}(t_{ij};\boldsymbol{\beta}_1,\ldots, \boldsymbol{\beta}_Q, \boldsymbol{B}_i)$ is a vector of length $Q$ and the $q$-th element represents the mean function of time $t_{ij}$ for the $q$-th biomarker characterized by fixed effects $\boldsymbol{\beta}_q=(\beta_{q1}, \ldots, \beta_{q,m_q})^\top$ of dimension $m_q$, and individual-specific random effects $\boldsymbol{b}_{iq}$.

For a set of measurements $\boldsymbol{X}_{ij}$ at time $t_{ij}$ for subject $i$, two scenarios may occur. In the first scenario, all biomarker values are fully observed and above the detection limits. Then no missing data is involved and the regression model in \eqref{eq:3.2} is sufficient. In the other scenario, some biomarker values fall below the detection limits and hence require imputation. Let $\boldsymbol{X}_{O,ij}$ of length $Q_O$ be the subset of $\boldsymbol{X}_{ij}$ and represent the vector of fully-observed biomarkers. Let $\boldsymbol{X}_{C,ij}$ be the complement of $\boldsymbol{X}_{O,ij}$ and represent the vector of left-censored biomarkers. Let $\boldsymbol{L}^\prime$ be the subset of $\boldsymbol{L}$ and denote the detection limits of biomarkers in $\boldsymbol{X}_{C,ij}$. Now we can represent $\boldsymbol{X}_{ij}\sim \mathcal{N}_Q(\boldsymbol{\mu}_{ij}, 
\boldsymbol{S}_i)$ as block-wise mean and covariance $\begin{bmatrix} \boldsymbol{X}_{O,ij} \\ \boldsymbol{X}_{C,ij}\end{bmatrix}\sim \mathcal{N}_Q\left(\begin{bmatrix} \boldsymbol{\mu}_{O,ij} \\
\boldsymbol{\mu}_{C,ij}\end{bmatrix}, \begin{bmatrix}
    \boldsymbol{S}_{O,i} & \boldsymbol{S}_{OC,i} \\
    \boldsymbol{S}_{CO,i} & \boldsymbol{S}_{C,i}
\end{bmatrix}\right)$, where $\boldsymbol{\mu}_{O,ij}$ and $\boldsymbol{\mu}_{C,ij}$ are the mean vectors of $\boldsymbol{X}_{O,ij}$ and $\boldsymbol{X}_{C,ij}$, respectively, with $\boldsymbol{S}_{O,i}$, $\boldsymbol{S}_{OC,i}$, $\boldsymbol{S}_{C,i}$ representing the variance-covariance matrix of $\boldsymbol{X}_{O,ij}$, the covariance matrix between $\boldsymbol{X}_{O,ij}$ and $\boldsymbol{X}_{C,ij}$ and the variance-covariance matrix of $\boldsymbol{X}_{C,ij}$, respectively. For the fully-observed $\boldsymbol{X}_{O,ij}$, the model remains $\boldsymbol{X}_{O,ij}\sim \mathcal{N}_{Q_O}(\boldsymbol{\mu}_{O,ij}, \boldsymbol{S}_{O,i})$. For the left-censored biomarkers $\boldsymbol{X}_{C,ij}$, we need to draw imputed values and also restrict the imputed values to be less than the detection limits. Thus, we obtain the imputed values from a truncated multivariate Gaussian distribution conditional on the fully-observed biomarker values as $\boldsymbol{X}_{C,ij}^{\text{imp}}|\boldsymbol{X}_{O,ij}\sim \mathcal{N}_{Q-Q_O}(\boldsymbol{\mu}_{C,ij}+\boldsymbol{S}_{CO,i}\boldsymbol{S}_{O,i}^{-1}(\boldsymbol{X}_{O,ij}-\boldsymbol{\mu}_{O,ij}), \boldsymbol{S}_{C,i}-\boldsymbol{S}_{CO,i}\boldsymbol{S}_{O,i}^{-1}\boldsymbol{S}_{OC,i}; \boldsymbol{X}_{C,ij}^{\text{imp}}<\boldsymbol{L}^\prime)$.

\noindent
\textbf{Priors for mean components}:
\begin{equation}
\begin{gathered}
    \beta_{qk}\sim \mathcal{N}(0, \lambda^2), \ k=1,\ldots,m_q, \text{ independently for }q=1,\ldots,Q;  \\
    \boldsymbol{\Sigma}_q=\text{Diag}(\sigma_{q1},
    \ldots,\sigma_{qP})\times \boldsymbol{\Omega}_q\times \text{Diag}(\sigma_{q1},
    \ldots,\sigma_{qP}), \ q=1,\ldots,Q;  \\
    \sigma_{qp}\sim \text{Half-Cauchy}(0,\tau_b), \ p = 1,\ldots,P, \ \boldsymbol{\Omega}_q\sim \text{LKJ}(\kappa_b), \text{ independently for }q=1,\ldots,Q,
\end{gathered}\label{eq:3.3}
\end{equation}
where we decompose $\boldsymbol{\Sigma}_q$ into a diagonal matrix of standard deviation $\sigma_{q1},\ldots,\sigma_{qP}$ for the random effects $\boldsymbol{b}_{iq}$ and a correlation matrix $\boldsymbol{\Omega}_q$. As recommended by \citet{gelman2006prior}, a Half-Cauchy prior is a reasonable starting point for small $\sigma_{qp}$ values since the conjugate Inverse-Gamma prior can be overly sensitive to the choice of hyper-parameters and may not act as an uninformative prior. The commonly-used conjugate Inverse-Wishart prior on the variance-covariance matrix has the similar problem, which is why we choose to use separate priors for the standard deviations and the correlation matrix. The Lewandowski-Kurowicka-Joe (LKJ) distribution is designed for positive definite correlation matrices and takes in one parameter $\kappa_b>0$, which tunes the strength of the correlations \citep{lewandowski2009generating}. When $\kappa_b=1$, the prior behaves like a uniform prior, giving every correlation matrix an equal probability. $\kappa_b>1$ favors correlation matrices with smaller correlations and the opposite applies when $\kappa_b<1$. For the hyper-parameters, we set $\lambda=100$, $\tau_b=2.5$ and $\kappa_b$=1. 

\noindent
\textbf{Priors for variance-covariance components}:
\begin{equation}
\begin{gathered}
    \boldsymbol{S}_i=\text{Diag}(s_{i1},\ldots,s_{iQ})\times \boldsymbol{D}_i\times \text{Diag}(s_{i1},\ldots,s_{iQ}), \text{ independently for }i=1,
    \ldots,N; \\
    \log(s_{iq}^2)\sim \mathcal{N}(\gamma_q, \Psi_q^2);  \\
    \gamma_q\sim \mathcal{N}(0, \phi^2), \ \Psi_q\sim \text{Half-Cauchy}(0, \zeta), \text{ independently for }q=1,\ldots,Q, \ i=1,\ldots,N,
\end{gathered}\label{eq3.4}
\end{equation}
where we use a similar strategy to that used for the variance structure of the random effects to decompose $\boldsymbol{S}_i$. Let $s_{iq}$ denote the residual standard deviation of the $q$-th biomarker for subject $i$ and $\boldsymbol{D}_i$ be the subject-specific correlation matrix between biomarkers. When $Q=2$, no restriction is needed for the off-diagonal correlation $r_{i}$ to maintain the positive-definiteness of $\boldsymbol{D}_i$. The prior for $r_{i}$ is specified as $(r_i+1)/2\sim \text{Beta}(a,b), a\sim \text{Exp}(\nu_a), b\sim \text{Exp}(\nu_b)$ and we set $\nu_a=\nu_b=0.1$. For the more general case with $Q>3$, we can re-parameterize the Cholesky factors of the correlation matrices in terms of hyperspherical coordinates or angles where the angles vary freely in the range of $[0,\pi)$ and specify priors accordingly \citep{ghosh2021bayesian, chen2024}.

\subsubsection{Component 2: Threshold Regression Model}
\label{sec3.2.2}
As we introduced in Section \ref{sec:threg}, we are interested in modeling the latent initial status $y_{0i}>0$ and drift $\zeta_i$ and fix $\sigma^2=1$. The two latent components are assumed to be associated with the individual-specific random effects and variances and baseline covariates through the following regression models:
\begin{equation}
\begin{gathered}
    \boldsymbol{W}_i=(1,\boldsymbol{b}^{\prime\top}_{i1},\ldots,\boldsymbol{b}^{\prime\top}_{iQ}, s^\prime_{i1},\ldots,s^\prime_{iQ}, \boldsymbol{Z}_i^\top)^\top;  \\
    b_{iqp}^\prime = \frac{b_{iqp}-0}{\sigma_{qp}}, \ p=1,\ldots,P, \ s_{iq}^\prime = \frac{\log(s_{iq}^2)-\gamma_q}{\Psi_q}, \ q=1,\ldots,Q; \\
    \log(y_{0i})=\boldsymbol{\alpha}^\top\boldsymbol{W}_i, \ \zeta_i=\boldsymbol{\eta}^\top\boldsymbol{W}_i,
\end{gathered}\label{eq3.5}
\end{equation}
where we standardize predictors from the longitudinal submodel to obtain stable estimates. Although we do not include the between-biomarker correlations  into the regression model as we believe the mean and variability are of more importance and more easily interpretable, such an extension is possible. We assume an independent $\mathcal{N}(0, \omega^2)$ prior for each regression coefficient and set $\omega=100$.

When modeling time-to-event outcomes with right censoring, one common approach is to use the survival probability at the censoring time as individual $i$'s contribution to the likelihood. However, this approach tends to produce biased estimates and attenuate the estimates towards the null in our simulation study. Therefore, we adopt the method by \citet{moghaddam2022bayesian} and draw imputations of the event times for the censored individuals from the parametric distribution in use, conditioned on the fact that the imputed event times should happen after the censoring times.

\subsection{Posterior Inference}
\label{sec3.3}
Since our model uses several non-conjugate priors, the posterior distribution is not available in closed form. Therefore, we use Hamiltonian Monte Carlo to draw posterior samples and implement our approach using {\textsf{Stan}} and the {\textsf{rstan}} package \citep{rstan} in \textsf{R}. For both the simulation and application, we run three chains with 15,000 iterations each, discarding the first 7,500 as burn-in. Convergence is assessed by visual inspection of the traceplots and the Gelman-Rubin $\hat{R}$ measure \citep{gelman2013bayesian}, showing maximum values below 1.1 and thus good convergence.

\section{Simulation}
\label{sec:simulation}

Before applying our proposed joint model to the motivating application, we conduct simulation studies with two primary aims: (1) to test the performance of the proposed method in consistently producing accurate estimates; and (2) to compare our method with commonly used two-stage approaches and demonstrate its superiority. Let $\theta_0$ denote the true value of parameter $\theta$ and $\hat{\theta}_r$ represent the posterior mean estimated from the $r$-th simulation, $r=1,\ldots,R$. The evaluation is conducted based on four criteria: (1) Coverage Rate: $1/R\sum^{R}_{r=1}\mathbbm{1}(\theta_0\in [L_r, U_r])$, where $L_r$ and $U_r$ represent the 2.5\% and 97.5\% percentiles of the posterior draws from the $r$-th simulation; (2) Bias: $1/R\sum_{r=1}^R(\hat{\theta}_r-\theta_0)^2$; (3) Average Interval Length: $1/R\sum_{r=1}^R(U_r-L_r)$; (4) Root Mean Squared Error (RMSE): $\sqrt{1/R\sum_{r=1}^R(\hat{\theta}_r-\theta_0)^2}$. For the simulations conducted below, we generate longitudinal biomarker trajectories for $N=1,000$ individuals, while the number of measurements per individual per biomarker is uniformly distributed between 6 to 15.

A simulation study with a single biomarker subject to a detection limit is presented in Section 1 of the Supplementary Materials. Here, we present the simulation analysis of two biomarkers ($Q=2$), with one biomarker subject to a detection limit. Longitudinal biomarker trajectories are generated based on the following regression model and parameter values:
\begin{equation}
\begin{gathered}
X_{ijq}=\sum_{k=1}^{m_q}\beta_{qk}t_{ij}^{k-1}+b_{iq1}+b_{iq2}t_{ij}+\epsilon_{ijq}, \ q=1,2, \ j=1,\ldots n_i, \ \text{independently for }i=1,\ldots,N;\\
    m_1=4 \text{ and } \boldsymbol{\beta}_1=(3.0, -0.2, 0.04, -0.001)^\top, \ m_2=3 \text{ and } \boldsymbol{\beta}_2=(6.6, 0.03, -0.05)^\top; \\
    \boldsymbol{b}_{i1}\sim \mathcal{N}_2(\boldsymbol{0},\boldsymbol{\Sigma}_1), \ \boldsymbol{\Sigma}_1=\text{Diag}(0.23, 0.05)\times \begin{pmatrix} 1 & 0.3 \\ 0.3 & 1 \end{pmatrix}\times \text{Diag}(0.23, 0.05); \\ 
    \boldsymbol{b}_{i2}\sim \mathcal{N}_2(\boldsymbol{0},\boldsymbol{\Sigma}_2), \ \boldsymbol{\Sigma}_2=\text{Diag}(0.65, 0.25)\times \begin{pmatrix} 1 & 0.18 \\ 0.18 & 1 \end{pmatrix}\times \text{Diag}(0.65, 0.25); \\
    \log(s_{i1}^2)\sim \mathcal{N}(-0.95, 0.45^2), \ \log(s_{i2}^2)\sim \mathcal{N}(0.50, 1^2), \ \frac{r_i+1}{2}\sim \text{Beta}(5.3, 12).
\end{gathered}\label{eq:3.6}
\end{equation}
and $X_{ij2}$ is subject to a detection limit with the limit of detection set to -1, and any simulated $X_{ij2}$ with values below -1 are treated as left-censored and require imputation. We perform $R=200$ simulations, with the average below-detection-limit rate of the second biomarker at 18.97\%, which mimics data in the application. Figure S2 in the Supplementary Materials shows a sample of the simulated biomarkers.

We then generate time-to-event outcomes $T_i$ based on random effects $b_{iq1}$ and $b_{iq2}$ and residual variability $s_{iq}^2$ simulated from the longitudinal submodel as follows:
\begin{equation}
\begin{gathered}
    b_{iqp}^\prime = \frac{b_{iqp}-0}{\sigma_{qp}}, \ p=1,2, \ s_{iq}^\prime = \frac{\log(s^2_{iq})-\gamma_q}{\Psi_q}, \ q=1,2;  \\
    \log(y_{0i})=3.7-0.2b_{i11}^\prime+0.15b_{i12}^\prime-0.04s_{i1}^\prime+0.1b_{i21}^\prime+0.25b_{i22}^\prime-0.1s_{i2}^\prime;  \\ \zeta_i=-3.8+0.85b_{i11}^\prime-0.9b_{i12}^\prime-0.02s_{i1}^\prime-0.28b_{i21}^\prime-0.36b_{i22}^\prime-0.2s_{i2}^\prime;  \\
    T_i\sim \text{Inverse-Gaussian}(-\frac{y_{0i}}{\zeta_i}, y_{0i}^2).
\end{gathered}\label{eq3.7}
\end{equation}
and we generate the right-censoring times $C_i$ from an independent exponential distribution. The average right-censoring rate across the 200 simulations is 9.7\%, which again mimics the data in the application.

\subsection{Alternative Methods}
\label{sec:4.1}

We briefly introduce two alternative methods based on a two-stage modeling strategy to be compared against our proposed joint model.

\textbf{Two-Stage Mixed Effects Model and Threshold Regression (TSMEThR)}  In the first two-stage alternative approach, we separately model the longitudinal biomarker trajectories and the time-to-event outcomes. In the first stage, we fit two separate linear mixed effects models for the two biomarkers as 
\begin{equation}
\begin{gathered}
    X_{ijq}=\sum_{k=1}^{m_q}\beta_{qk}t_{ij}^{k-1}+b_{iq1}+b_{iq2}t_{ij}+\epsilon_{ijq}, \ q=1,2, \ j=1,\ldots,n_i,\\
    \text{independently for } i=1,\ldots,N.
\end{gathered}\label{eq:3.8}
\end{equation}
For $X_{ij1}$, we fit a regular linear mixed effects model with \textsf{lme()} from the \textsf{nlme} package. For $X_{ij2}$,
to account for the detection limit and left-censored values, we fit the model using \textsf{mixed\_model()} from the \textsf{GLMMadaptive} package, specifying the family argument as ``\textsf{censored.normal}", which is designed to handle mixed effects models with censored and continuous normally distributed outcomes. For each individual $i$, estimates of random effects $\hat{b}_{iq1}$ and $\hat{b}_{iq2}$ are obtained from the model output, and the residual variance for individual $i$ is estimated by $\hat{s}_{iq}^2=\sum_{j=1}^{n_i}\hat{\epsilon}_{ijq}^2$.

In the second stage, we model the time-to-event outcome by Bayesian threshold regression, following the model and prior specifications stated in Section \ref{sec3.2.2}. The only difference is that, here we standardize the predictors obtained from the first stage based on empirical means and standard deviations. Standardized random effects are calculated by $\hat{b}_{iqp}^\prime=(\hat{b}_{iqp}-\bar{\hat{b}}_{qp})/\hat{\sigma}_{qp}$ for $p=1,2,q=1,2$, where $\bar{\hat{b}}_{qp}=1/N\sum_{i=1}^N\hat{b}_{iqp}$ and $\hat{\sigma}_{qp}=\sqrt{1/(N-1)\sum_{i=1}^N(\hat{b}_{iqp}-\bar{\hat{b}}_{qp})^2}$. The standardized residual variances are calculated similarly as $\hat{s}_{iq}^\prime=(\log(\hat{s}_{iq}^2)-\bar{\hat{s}}_q)/\hat{\sigma}_{sq}$ with $\bar{\hat{s}}_q=1/N\sum_{i=1}^N\log(\hat{s}_{iq}^2)$ and $\hat{\sigma}_{sq}=\sqrt{1/(N-1)\sum_{i=1}^N(\log(\hat{s}_{iq}^2)-\bar{\hat{s}}_q)^2}$. 

\textbf{Two-Stage Individual Model and Threshold Regression (TSIMThR)}  In the second alternative method, we aim to demonstrate that moving from a two-stage approach to joint modeling while maintaining the exact same model structure can lead to substantial improvement in performance. Here, we first fit the longitudinal biomarker measurements using the model and prior specified in Section \ref{sec3.2.1} and obtain posterior mean estimates. Then we regress the time-to-event outcome on these posterior means based on the model and prior specification stated in Section \ref{sec3.2.2}. The only difference from the joint model is that we break the joint model into two separate components and connect them through the posterior mean estimates from the first-stage model.

\subsection{Results}
\label{sec:4.2}

\begin{figure}
\begin{center}
\includegraphics[width=7in]{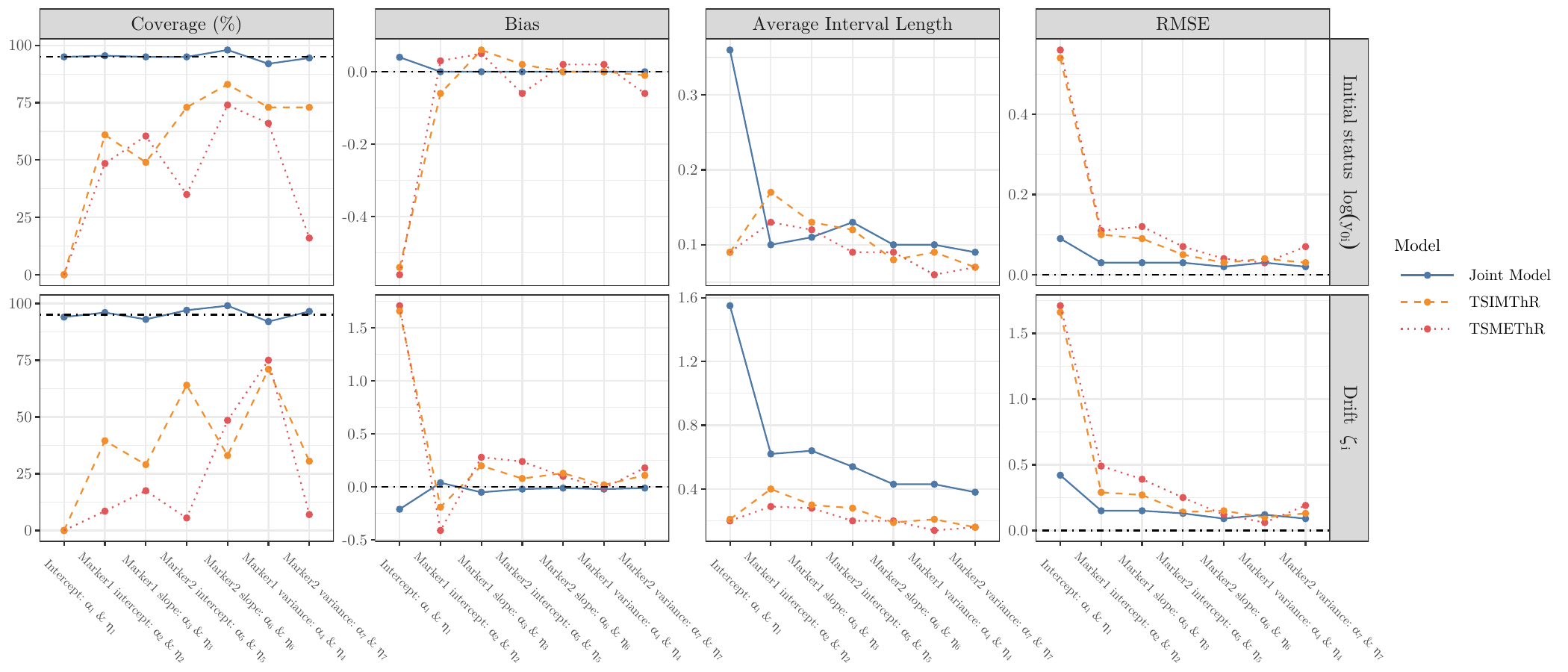}
\end{center}
\caption{\small Visualization of the simulation results when $Q=2$ with one biomarker subject to a detection limit. The dot-dashed horizontal line in each panel indicates reference values: 95\% for Coverage (\%), 0 for Bias and RMSE. \label{fig:Q2_result}}
\end{figure}

Table
\ref{tab:Q2_alpha}, Table \ref{tab:Q2_eta} and Figure \ref{fig:Q2_result} compare our proposed joint model with two-stage approaches when $Q=2$ with one biomarker subject to a detection limit. Simulation results for all other parameters in the joint model can be found in Table S3 in the Supplementary Materials. The joint model is able to consistently produce accurate estimates with high coverage rates, whereas the two alternative methods exhibit greater bias and poorer coverage, especially for the intercept terms. The coverage and bias of estimates from TSIMThR have significantly improved compared to TSMEThR; however, TSIMThR still has difficulty producing accurate estimates, especially for coefficients related to the residual variability. Comparisons of the joint model with TSIMThR highlight that, despite using the same model framework for the longitudinal and survival components, moving from two-stage to joint modeling substantially reduces bias and improves the accuracy and coverage of the estimates. Overall, these simulation studies demonstrate the effectiveness and superiority of the proposed joint model.

\begin{table}
\caption{\small Simulation results including coverage rate, bias, average interval length and RMSE of $\boldsymbol{\alpha}$ (regression coefficients of $\log(y_{0i})$) and comparing the joint model with alternative two-stage approaches when $Q=2$ with one biomarker subject to a detection limit.  \label{tab:Q2_alpha}}
\renewcommand{\arraystretch}{1}
\begin{center}
\begin{tabular}{lrrrrr}
\hline
True Values & Model & Coverage (\%) & Bias & Average Interval Length & RMSE\\\hline
$\alpha_1$ = 3.7 & \textbf{Joint Model} & 95.0 & 0.04 & 0.36 & 0.09\\ 
      & TSMEThR & 0.0 & -0.56 & 0.09 & 0.56\\
      & TSIMThR & 0.0 & -0.54 & 0.09 & 0.54 \\
      $\alpha_2$ = -0.2 & \textbf{Joint Model} & 95.5 & -0.00 & 0.10 & 0.03\\
      & TSMEThR & 48.5 & 0.03 & 0.13 & 0.11\\
      & TSIMThR & 62.5 & -0.06 & 0.17 & 0.11 \\
    $\alpha_3$ = 0.15 & \textbf{Joint Model} & 95.0 & 0.00 & 0.11 & 0.03\\
      & TSMEThR & 60.5 & 0.05 & 0.12 & 0.12\\
      & TSIMThR & 50.0 & 0.07 & 0.13 & 0.10 \\
      $\alpha_4$ = -0.04 & \textbf{Joint Model} & 92.0 & 0.00 & 0.10 & 0.03\\
      & TSMEThR & 66.0 & 0.02 & 0.06 & 0.03\\
      & TSIMThR & 75.5 & 0.00 & 0.09 & 0.04 \\
      $\alpha_5$ = 0.1 & \textbf{Joint Model} & 95.0 & 0.00 & 0.13 & 0.03\\
      & TSMEThR & 35.0 & -0.06 & 0.09 & 0.07\\
      & TSIMThR & 79.0 & 0.01 & 0.12 & 0.05 \\
      $\alpha_6$ = 0.25 & \textbf{Joint Model} & 98.0 & 0.00 & 0.10 & 0.02\\
      & TSMEThR & 74.0 & 0.02 & 0.09 & 0.04\\
      & TSIMThR & 81.5 & 0.00 & 0.08 & 0.03 \\
      $\alpha_7$ = -0.1 & \textbf{Joint Model} & 94.5 & 0.00 & 0.09 & 0.02\\
      & TSMEThR & 16.0 & -0.06 & 0.07 & 0.07\\
      & TSIMThR & 74.5 & -0.02 & 0.07 & 0.03 \\
      \hline
\end{tabular}
\end{center}
\end{table}

\begin{table}
\caption{\small Simulation results including coverage rate, bias, average interval length and RMSE of $\boldsymbol{\eta}$ (regression coefficients of $\zeta_i$) and comparing the joint model with alternative two-stage approaches when $Q=2$ with one biomarker subject to a detection limit.  \label{tab:Q2_eta}}
\renewcommand{\arraystretch}{1}
\begin{center}
\begin{tabular}{lrrrrr}
\hline
True Values & Model & Coverage (\%) & Bias & Average Interval Length & RMSE\\\hline
$\eta_1$ = -3.8 & \textbf{Joint Model} & 94.0 & -0.21 & 1.55 & 0.42\\ 
      & TSMEThR & 0.0 & 1.71 & 0.20 & 1.71\\
      & TSIMThR & 0.0 & 1.66 & 0.21 & 1.66 \\
      $\eta_2$ = 0.85 & \textbf{Joint Model} & 96.0 & 0.04 & 0.62 & 0.15\\
      & TSMEThR & 8.5 & -0.41 & 0.29 & 0.49\\
      & TSIMThR & 39.5 & -0.19 & 0.40 & 0.29 \\
    $\eta_3$ = -0.9 & \textbf{Joint Model} & 93.0 & -0.05 & 0.64 & 0.15\\
      & TSMEThR & 17.5 & 0.28 & 0.28 & 0.39\\
      & TSIMThR & 29.0 & 0.20 & 0.30 & 0.27 \\
      $\eta_4$ = -0.02 & \textbf{Joint Model} & 92.0 & -0.02 & 0.43 & 0.12\\
      & TSMEThR & 75.0 & -0.01 & 0.14 & 0.06\\
      & TSIMThR & 71.0 & 0.02 & 0.21 & 0.10 \\
      $\eta_5$ = -0.28 & \textbf{Joint Model} & 97.0 & -0.02 & 0.54 & 0.13\\
      & TSMEThR & 5.5 & 0.24 & 0.20 & 0.25\\
      & TSIMThR & 64.0 & 0.08 & 0.28 & 0.14 \\
      $\eta_6$ = -0.36 & \textbf{Joint Model} & 99.0 & -0.01 & 0.43 & 0.09\\
      & TSMEThR & 48.5 & 0.10 & 0.20 & 0.12\\
      & TSIMThR & 33.0 & 0.13 & 0.19 & 0.15 \\
      $\eta_7$ = -0.2 & \textbf{Joint Model} & 96.5 & -0.01 & 0.38 & 0.09\\
      & TSMEThR & 7.0 & 0.18 & 0.16 & 0.19\\
      & TSIMThR & 30.5 & 0.11 & 0.16 & 0.13 \\
      \hline
\end{tabular}
\end{center}
\end{table}

\section{Application to the Study of Women's Health Across the Nation (SWAN)}
\label{sec:application}

\subsection{Study Description}
\label{sec:5.1}
The Study of Women's Health Across the Nation (SWAN) is a multi-site, multi-racial and ethnic longitudinal cohort study of the menopausal transition \citep{article}. In 1996, 3,302 women aged 42-52 who self-identified as one of the predesignated racial or ethnic groups were recruited for the SWAN longitudinal study from seven sites. Eligibility criterion for SWAN included having a uterus and at least one ovary, having a menstrual period in the past three months, no use of exogenous hormones in the past three months, and not pregnant or lactating. Participants had their blood drawn, physical measures collected and completed questionnaires at baseline and during 16 annual follow-up visits. FSH (mIU/ml) hormone levels were assayed at each visit while a subset of 1,536 women had their AMH (pg/ml) hormone levels assayed at up to 12 visits using banked serum. Age at FMP was determined retrospectively based upon information regarding menstrual bleeding patterns obtained during each follow-up visit. The FMP was defined retrospectively as the bleeding episode preceding 12 months of amenorrhea. 

\begin{figure}
\begin{center}
\includegraphics[width=5in]{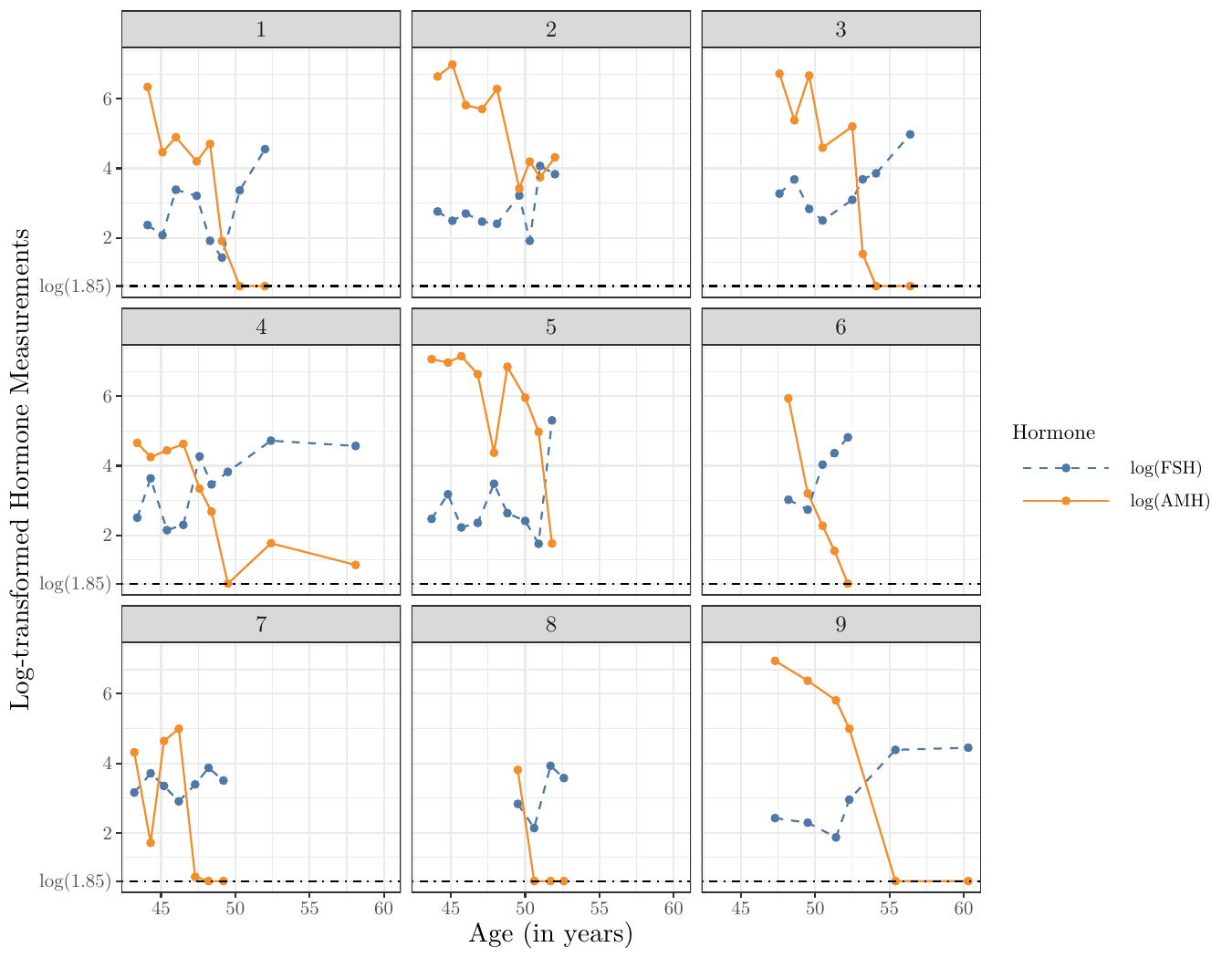}
\end{center}
\caption{\small Hormone trajectories by age for nine randomly selected participants. The dot-dashed horizontal line in each panel represents the log-transformed detection limit for AMH at $\log(1.85)$. 
 \label{fig:data_sample}}
\end{figure}

After excluding women a) with fewer than four AMH or FSH measurements, and/or b) missing baseline body mass index (BMI) and excluding measurements taken when the female was reported to be on hormone therapy, our final analytical sample contains 976 women with 6,295 visits. Among them, 917 women had documented age at FMP ranging from age 45.6 to 62.1, while for the remaining 59, we treat age at their last visit with an observed hormone value as the censoring age. The assay used to measure AMH has a detection limit of 1.85 pg/ml and 24.9\% of AMH measurements in our analytical sample fall below this threshold. We use log-transformed FSH and AMH values in our subsequent analysis to reduce the skewness of the data. Figure \ref{fig:data_sample} displays longitudinal FSH and AMH measurements for nine randomly selected women. We observe that FSH increases during the menopausal transition while AMH decreases.

\subsection{Model for SWAN Data} 
\label{sec:5.2}
The time variable for modeling the longitudinal hormone trajectories is the shifted age at each visit (age-42 years, the youngest observed age at baseline) and the regression model is as follows:
\begin{equation}
\begin{gathered}
\text{age}_{\text{shift}}=\text{age at each visit}-42; \\
    \log(\text{FSH}_{ij})=\beta_{11}+\beta_{12}\text{age}_{\text{shift},ij}+\beta_{13}\text{age}_{\text{shift},ij}^2+\beta_{14}\text{age}_{\text{shift},ij}^3+b_{i11}+b_{i12}\text{age}_{\text{shift},ij}+\epsilon_{ij1};  \\
    \log(\text{AMH}_{ij})=\beta_{21}+\beta_{22}\text{age}_{\text{shift},ij}+\beta_{23}\text{age}_{\text{shift},ij}^2+b_{i21}+b_{i22}\text{age}_{\text{shift},ij}+\epsilon_{ij2}.
\end{gathered}\label{eq:3.9}
\end{equation}
and AMH values below the detection limit are imputed as in Section \ref{sec3.2.1}. 

To maintain consistency with the time variable used for modeling hormones, we also shift age at FMP, subtracting 42 years from the observed age at FMP. Then we regress the two latent components of threshold regression on the hormone parameters and baseline BMI: 
\begin{equation}
    \begin{gathered}
    b_{iqp}^\prime=\frac{b_{iqp}-0}{\sigma_{qp}}, \ p=1,2, \ s_{iq}^\prime=\frac{\log(s_{iq}^2)-\gamma_q}{\Psi_b}, \ q=1,2; \\
    \log(y_{0i})=\alpha_1+\alpha_2b_{i11}^\prime+\alpha_3b_{i12}^\prime+\alpha_4s_{i1}^\prime+\alpha_5b_{i21}^\prime+\alpha_6b_{i22}^\prime+\alpha_7s_{i2}^\prime+\alpha_8\text{BMI}_i;  \\
    \zeta_i=\eta_1+\eta_2b_{i11}^\prime+\eta_3b_{i12}^\prime+\eta_4s_{i1}^\prime+\eta_5b_{i21}^\prime+\eta_6b_{i22}^\prime+\eta_7s_{i2}^\prime+\eta_8\text{BMI}_i.
\end{gathered}\label{eq:3.10}
\end{equation}

\subsection{Results} 
\label{sec:5.3}

\begin{figure}
\begin{center}
\includegraphics[width=6in]{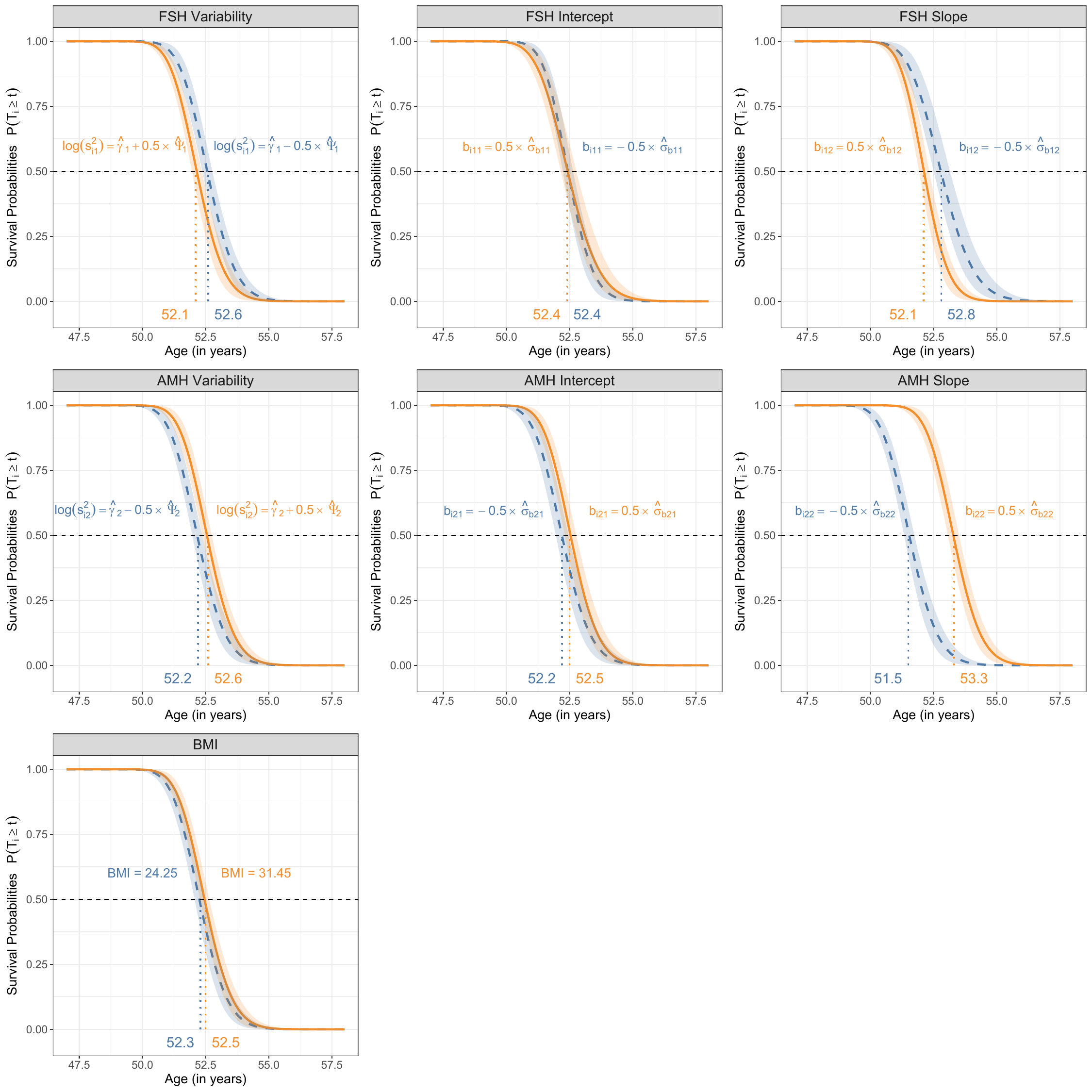}
\end{center}
\caption{\small Estimated survival curves for an average individual at different levels of hormone parameters and BMI. In each graph, we vary one covariate and fix the others at their mean values. In the top panel, we present the survival curves when varying log-transformed FSH variability, FSH intercept and slope by one-half standard deviation around the mean, respectively. The middle panel presents the similar survival curves corresponding to AMH. The bottom panel shows the survival curves and median age at FMP when varying BMI by half a standard deviation around the mean BMI of the study sample. 
 \label{fig:result_survcurv}}
\end{figure}

Because the individual latent components are less interpretable independently, we present posterior estimates of the model coefficients
in Table S4 and Table S5 in the Supplementary Materials. Instead, we focus on
estimating survival curves and the median age at FMP in Figure \ref{fig:result_survcurv} and Table \ref{tab:median_age} for an average individual at different levels of hormone parameters in order to better understand and illustrate how the hormone affects age at FMP. In each subplot, we vary one hormone parameter (e.g., FSH variability) and fix the other parameters at the mean values based on the posterior mean estimates (e.g, fix FSH slope at 0 and log-transformed AMH variability at the posterior mean of $\gamma_2=\hat{\gamma}_2$). As seen in the top panel, when a individual's log-transformed FSH variability is half a standard deviation above the mean ($\log(s_{i1}^2)=\hat{\gamma}_1+0.5\hat{\Psi}_1$), the survival curve declines more rapidly at younger ages and more slowly at older ages, compared to that of a individual with log-transformed FSH variability half a standard deviation below the mean ($\log(s_{i1}^2)=\hat{\gamma}_1-0.5\hat{\Psi}_1$), with largely non-overlapping 95\% CrIs and a moderate gap between the two curves. The median age at FMP for women with log-transformed FSH variability half a standard deviation above the mean ($\log(s_{i1}^2)=\hat{\gamma}_1+0.5\hat{\Psi}_1$) is 52.1 (95\% CrI: 52.0, 52.3) years, which is 0.5 (95\%: 0.1, 0.7) years earlier than those with log-transformed FSH variability half a standard deviation below the mean ($\log(s_{i1}^2)=\hat{\gamma}_1-0.5\hat{\Psi}_1$). However, when varying the FSH intercept by one standard deviation (comparing $b_{i11}=0.5\hat{\sigma}_{11}$ and $b_{i11}=-0.5\hat{\sigma}_{11}$), there is no notable differences in the survival curves or the median age at FMP. FSH slope is another strong predictor of age at FMP, with one standard deviation increase around the mean (comparing $b_{i12}=-0.5\hat{\sigma}_{12}$ and $b_{i12}=0.5\hat{\sigma}_{12}$) pushing the median age at FMP to be 0.7 (95\% CrI: 0.3, 1.1) years earlier.

\begin{table}
\caption{\small Posterior mean and 95\% credible intervals (CrI) for the estimated median age at FMP for an average individual at different levels of hormone parameters and BMI. We vary one covariate in each cell and fix the others at their mean values. The hormone parameter and BMI values correspond to those in Figure \ref{fig:result_survcurv}. \label{tab:median_age}}
\renewcommand{\arraystretch}{1}
\begin{center}
\begin{tabular}{lrr}
\hline
Covariate & Median Age at FMP & 95\% Credible Interval \\
\hline
\multicolumn{3}{l}{\textbf{FSH Variability}} \\
Low ($\hat{\gamma}_1 - 0.5\hat{\Psi}_1$) & 52.6 & (52.4, 52.8) \\
High ($\hat{\gamma}_1 + 0.5\hat{\Psi}_1$) & 52.1 & (52.0, 52.3) \\
\multicolumn{3}{l}{\textbf{FSH Intercept}} \\
Low ($-0.5\hat{\sigma}_{b11}$) & 52.4 & (52.2, 52.6) \\
High ($0.5\hat{\sigma}_{b11}$) & 52.4 & (52.2, 52.8) \\
\multicolumn{3}{l}{\textbf{FSH Slope}} \\
Low ($-0.5\hat{\sigma}_{b12}$) & 52.8 & (52.5, 53.1) \\
High ($0.5\hat{\sigma}_{b12}$) & 52.1 & (51.9, 52.3) \\
\multicolumn{3}{l}{\textbf{AMH Variability}} \\
Low ($\hat{\gamma}_2 - 0.5\hat{\Psi}_2$) & 52.2 & (52.0, 52.4) \\
High ($\hat{\gamma}_2 + 0.5\hat{\Psi}_2$) & 52.6 & (52.4, 52.7) \\
\multicolumn{3}{l}{\textbf{AMH Intercept}} \\
Low ($-0.5\hat{\sigma}_{b21}$) & 52.2 & (52.0, 52.5) \\
High ($0.5\hat{\sigma}_{b21}$) & 52.5 & (52.3, 52.8) \\
\multicolumn{3}{l}{\textbf{AMH Slope}} \\
Low ($-0.5\hat{\sigma}_{b22}$) & 51.5 & (51.3, 51.7) \\
High ($0.5\hat{\sigma}_{b22}$) & 53.3 & (53.1, 53.5) \\
\multicolumn{3}{l}{\textbf{BMI}} \\
Low (24.25) & 52.3 & (52.1, 52.4) \\
High (31.45) & 52.5 & (52.3, 52.6) \\
\hline
\end{tabular}
\end{center}
\end{table}

The middle panel of Figure \ref{fig:result_survcurv} presents the survival curves when we vary AMH parameters. Unlike FSH variabilities, higher AMH variability is associated with a later age at FMP. For individuals with log-transformed AMH variability half a standard deviation above the mean ($\log(s_{i2}^2)=\hat{\gamma}_2+0.5\hat{\Psi}_2$), the median age at FMP is 52.6 (95\% CrI: 52.4, 52.7) years, 0.4 (95\% CrI: 0.1, 0.7) years later than those with log-transformed AMH variability half a standard deviation below the mean ($\log(s_{i2}^2)=\hat{\gamma}_2-0.5\hat{\Psi}_2$). Varying the AMH intercept by one standard deviation exhibits a similar trend (comparing $b_{i21}=0.5\hat{\sigma}_{21}$ and $b_{i21}=-0.5\hat{\sigma}_{21}$), however, the 95\% CrIs overlap across all ages. AMH slope has the most significant impact among all measured parameters, and increasing AMH slope by one standard deviation around the mean (comparing $b_{i22}=-0.5\hat{\sigma}_{22}$ and $b_{i22}=0.5\hat{\sigma}_{22}$) leads to a large delay in the median age at FMP from 51.5 (95\% CrI: 51.3, 51.7) years to 53.3 (95\% CrI: 53.1, 53.5) years. 

In the bottom panel in Figure \ref{fig:result_survcurv}, we observe that women with higher BMI tend to experience FMP at a slightly later age, consistent with findings from previous studies \citep{zhu2018body}. Overall, the results suggest that greater variability and a faster rise of FSH is associated with an earlier age at FMP, while greater variability and a slower decline of AMH is associated with a later age at FMP, adjusting for the effects of each hormone simultaneously. The effect of BMI on age at FMP is small after adjusting for hormone parameters.

\section{Discussion}
\label{sec:discussion}
Our method addresses three long-standing limitations of the current joint modeling of longitudinal and survival data. First, we expand the model from a single biomarker to multiple biomarkers, including those with detection limits. Secondly, we incorporate biomarker variability as a predictor while retaining the role of the mean. Finally, the shift from classic Cox model to the flexible threshold regression relaxes the proportional hazards assumptions. This work confirmed previous work that indicated more rapid increases in FSH and more rapid decreases in AMH during menopausal transition are associated with earlier FMP, but also allow accurate estimation of this joint effect, showing that the effect of these hormones are to some degree independent of each other, and that the relative effect of AMH is stronger than the relative effect of FSH. Further, variabilities of both measures are important predictors, although the effect is in different directions: more variability in FSH is associated with earlier FMP, while more variability in AMH is associated with later FMP.

There are, of course, limitations to our approach. 
We did not consider time-varying covariates; \cite{lee2010} provides extensions of threshold regression that can accommodate time varying covariates.
Future research that will improve the utility of the model is to allow for time-varying covariates in the threshold regression to leverage additional time-varying information.  Other extensions include accommodating non-Gaussian distribute and discrete longitudinal predictors with respect to both mean trends and, e.g., overdispersion measures.

\section*{Acknowledgments}

The Study of Women's Health Across the Nation (SWAN) has grant support from the National Institutes of Health (NIH), DHHS, through the National Institute on Aging (NIA), the National Institute of Nursing Research (NINR) and the NIH Office of Research on Women’s Health (ORWH) (Grants U01NR004061; U01AG012505, U01AG012535, U01AG012531, U01AG012539, U01AG012546, U01AG012553, U01AG012554, U01AG012495, and U19AG063720). The content of this article is solely the responsibility of the authors and does not necessarily represent the official views of the NIA, NINR, ORWH or the NIH. This research was also partly supported through computational resources and services provided by Advanced Research Computing (ARC), a division of Information and Technology Services (ITS) at the University of Michigan, Ann Arbor.

Clinical Centers: \textit{University of Michigan, Ann Arbor---Carrie Karvonen-Gutierrez, PI 2021–present, Siobán Harlow, PI 2011–2021, MaryFran Sowers, PI 1994-2011; Massachusetts General Hospital, Boston, MA---Sherri‐Ann Burnett‐Bowie, PI 2020–Present; Joel Finkelstein, PI 1999–2020; Robert Neer, PI 1994–1999; Rush University, Rush University Medical Center, Chicago, IL---Imke Janssen, PI 2020–Present; Howard Kravitz, PI 2009–2020; Lynda Powell, PI 1994–2009; University of California, Davis/Kaiser---Elaine Waetjen and Monique Hedderson, PIs 2020–Present; Ellen Gold, PI 1994-2020; University of California, Los Angeles---Arun Karlamangla, PI 2020–Present; Gail Greendale, PI 1994-2020; Albert Einstein College of Medicine, Bronx, NY---Carol Derby, PI 2011–present, Rachel Wildman, PI 2010–2011; Nanette Santoro, PI 2004–2010; University of Medicine and Dentistry–New Jersey Medical School, Newark---Gerson Weiss, PI 1994–2004, and the University of Pittsburgh, Pittsburgh, PA---Rebecca Thurston, PI 2020–Present; Karen Matthews, PI 1994-2020}.

NIH Program Office: \textit{National Institute on Aging, Bethesda, MD---Rosaly Correa-de-Araujo 2020-present; Chhanda Dutta 2016-present; Winifred Rossi 2012–2016; Sherry Sherman 1994–2012; Marcia Ory 1994–2001; National Institute of Nursing Research, Bethesda, MD---Program Officers}.

Central Laboratory: \textit{University of Michigan, Ann Arbor---Daniel McConnell (Central Ligand Assay Satellite Services)}.

Coordinating Center: \textit{University of Pittsburgh, Pittsburgh, PA---Maria Mori Brooks, PI 2012-present; Kim Sutton-Tyrrell, PI 2001–2012; New England Research Institutes, Watertown, MA---Sonja McKinlay, PI 1995–2001}.

Steering Committee:	Susan Johnson, Current Chair, Chris Gallagher, Former Chair.

We thank the study staff at each site and all the women who participated in SWAN.

The authors would also like to thank the SWAN team at the University of Michigan for providing the datasets for analysis and Dr. John Randolph for his expertise in women's hormone development and for his review and comments of the manuscript.

\section*{Funding}

This work was supported in part by National Institute on Aging Grant 1-R56-AG066693.

\section*{Disclosure Statement}

The authors report there are no competing interests to declare.

\section*{Supplementary Materials}

The Supplementary Materials provide information on additional results for the simulations presented in the main text, extra simulations, posterior parameter estimates for the data application, and results of posterior predictive checking of the joint model for the application data.

\bibliographystyle{apalike}
\bibliography{references}  






\end{document}


\maketitle

\section{Simulation}
\label{sec:sim}

In this section, we present another set of simulations, the corresponding results and some additional simulation results of the simulation presented in the main text.

\subsection{Simulation 1: One biomarker, $Q=1$}\label{sec:1.1}

\begin{figure}
\begin{center}
\includegraphics[width=5in]{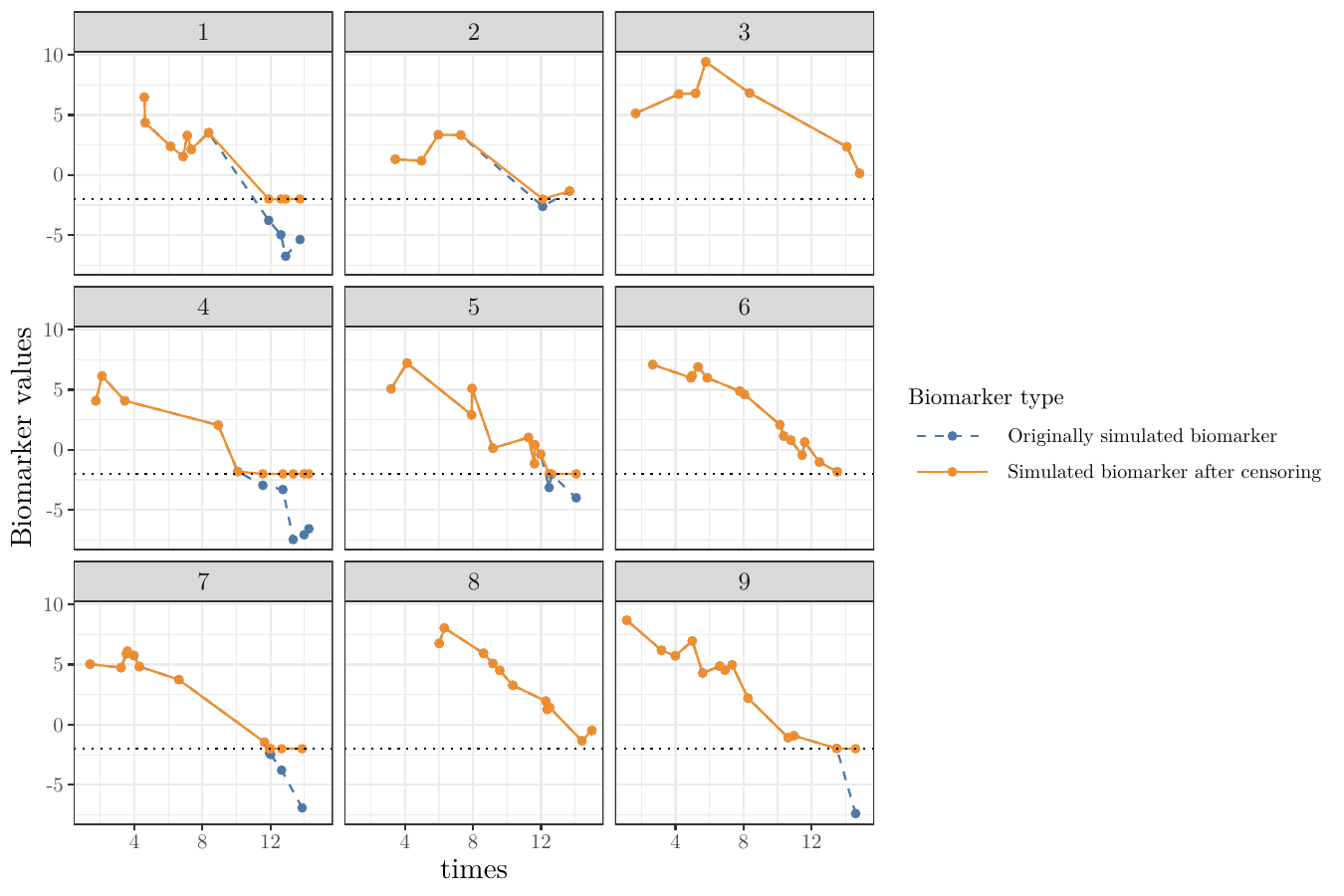}
\end{center}
\caption{\small Simulated biomarker values for a single biomarker with a detection limit for nine randomly selected individuals. The dashed lines represent the originally simulated values before applying the detection limit, and the solid lines represent the biomarker values after censoring. The dotted 
horizontal line in each panel represents the detection limit at -2. \label{fig:Q1_sim_plot}}
\end{figure}

In addition to the simulation presented in the main text, we also conduct simulations with a single biomarker subject to a detection limit. We assume that the fixed effects are up to a quadratic term of time, while for the random effects, we include only random intercept and slope. The longitudinal biomarker trajectories are simulated based on the following regression model and parameter values:
\begin{equation}
\begin{gathered}
    X_{ij1}=\beta_{11}+\beta_{12}t_{ij}+\beta_{13}t_{ij}^2+b_{i11}+b_{i12}t_{ij}+\epsilon_{ij1},\ j=1,\ldots,n_i, \ \text{independently for }i=1,\ldots,N; \\
    m_1=3 \text{ and } \boldsymbol{\beta}_1=(6.5, 0.07, -0.06)^\top; \\
    \boldsymbol{b}_{i1}\sim \mathcal{N}_2(\boldsymbol{0},\boldsymbol{\Sigma}_1), \ \boldsymbol{\Sigma}_1=\text{Diag}(0.75, 0.3)\times \begin{pmatrix}
        1 & -0.1 \\ -0.1 & 1
    \end{pmatrix}\times \text{Diag}(0.75, 0.3); \\
    \log(s_{i1}^2)\sim \mathcal{N}(0.45, 1^2).
\end{gathered}\label{eq:S3.1}
\end{equation}
The detection limit is set to -2, and any simulated biomarker values below -2 are treated as left-censored and require imputation. We perform $R=200$ simulations, with the average below-detection-limit rate at 19.25\%. A sample of simulated biomarkers is displayed in Figure \ref{fig:Q1_sim_plot}.

As the next step, we generate time-to-event outcomes $T_i$ based on random effects $b_{i11}$ and $b_{i12}$ along with the residual variability $s_{i1}^2$ simulated in the longitudinal submodel as
\begin{equation}
\begin{gathered}
    b_{i1p}^\prime = \frac{b_{i1p}-0}{\sigma_{1p}}, \ p=1,2, \ s_{i1}^\prime = \frac{\log(s_{i1}^2)-\gamma_1}{\Psi_1}; \\
    \log(y_{0i})=3.5+0.3b_{i11}^\prime+0.2b_{i12}^\prime+0.15s_{i1}^\prime, \ \zeta_i=-3-0.8b_{i11}^\prime-0.02b_{i12}^\prime-0.3s_{i1}^\prime;\\
    T_i\sim \text{Inverse-Gaussian}(-\frac{y_{0i}}{\zeta_i}, y_{0i}^2).
\end{gathered}\label{eq:S3.2}
\end{equation}
and we generate right-censoring times $C_i$ from an independent exponential distribution. The average censoring rate is 9.7\%.

To compare with the joint model, we adopt the same two alternative two-stage approaches described in Section 4.1 in the main text. The TSIMThR remains unchanged. For TSMEThR, now we only need to fit one set of linear mixed effects models using \textsf{mixed\_model()} as described in Section 4.1 in the main text to account for the detection limit issue. The rest of the steps remain the same.

\subsection{Simulation 2: Two biomarker, $Q=2$}
\label{sec:1.2}

\begin{figure}
\begin{center}
\includegraphics[width=5in]{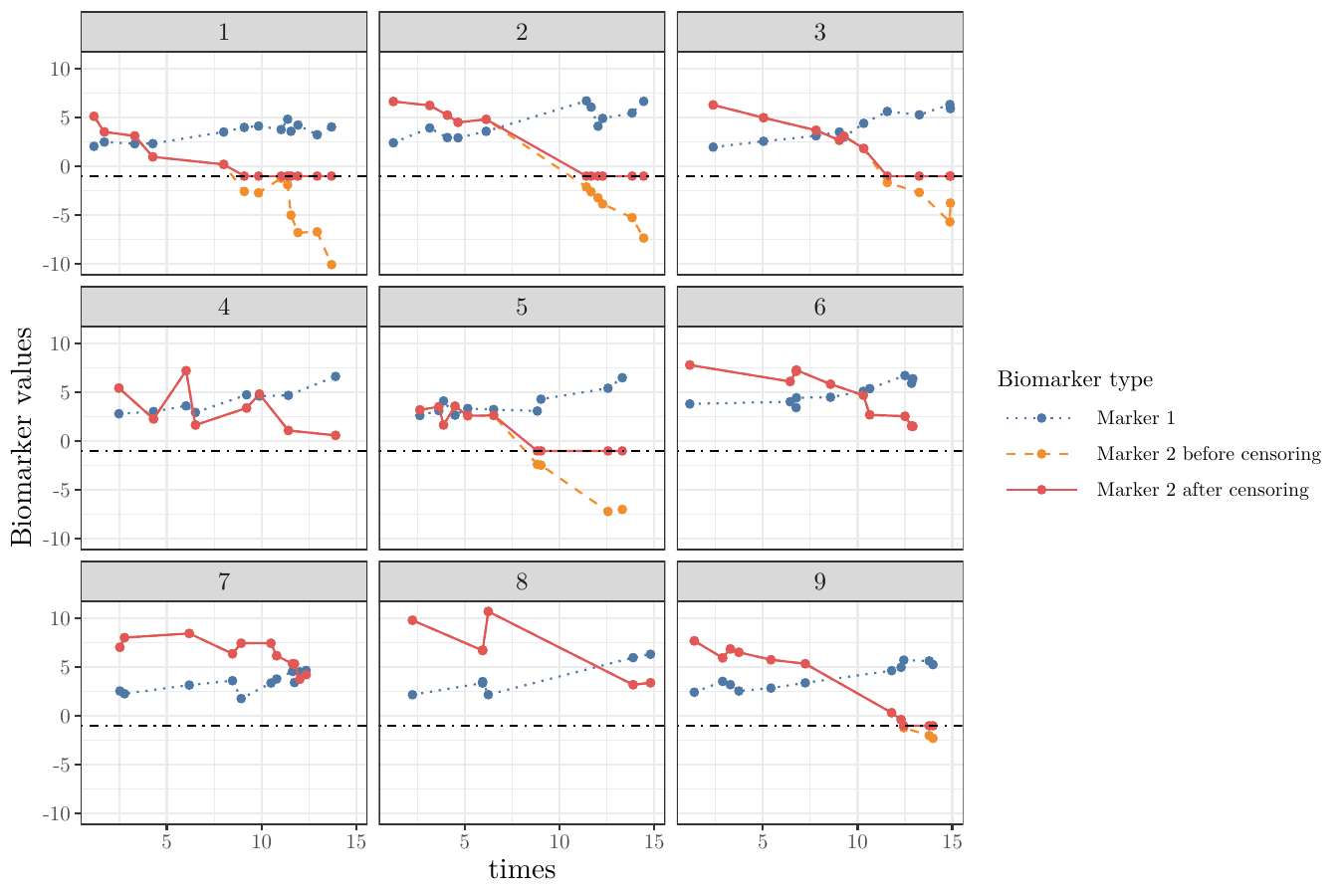}
\end{center}
\caption{\small Simulated biomarker values when $Q=2$ with one biomarker subject to a detection limit for nine randomly selected individuals. The dotted lines represent the simulated values for marker 1. The dashed lines represent the originally simulated values for marker 2 with the solid lines representing the simulated values for marker 2 after applying the detection limit. The dot-dashed horizontal line in each panel represents the detection limit at -1. \label{fig:Q2_sim_plot}}
\end{figure}

Following the description in the main text, a sample of the simulated biomarkers when $Q=2$ with one biomarker subject to a detection limit is shown in Figure \ref{fig:Q2_sim_plot}.

\subsection{Results}
\label{sec:1.3}

\begin{figure}
\begin{center}
\includegraphics[width=7in]{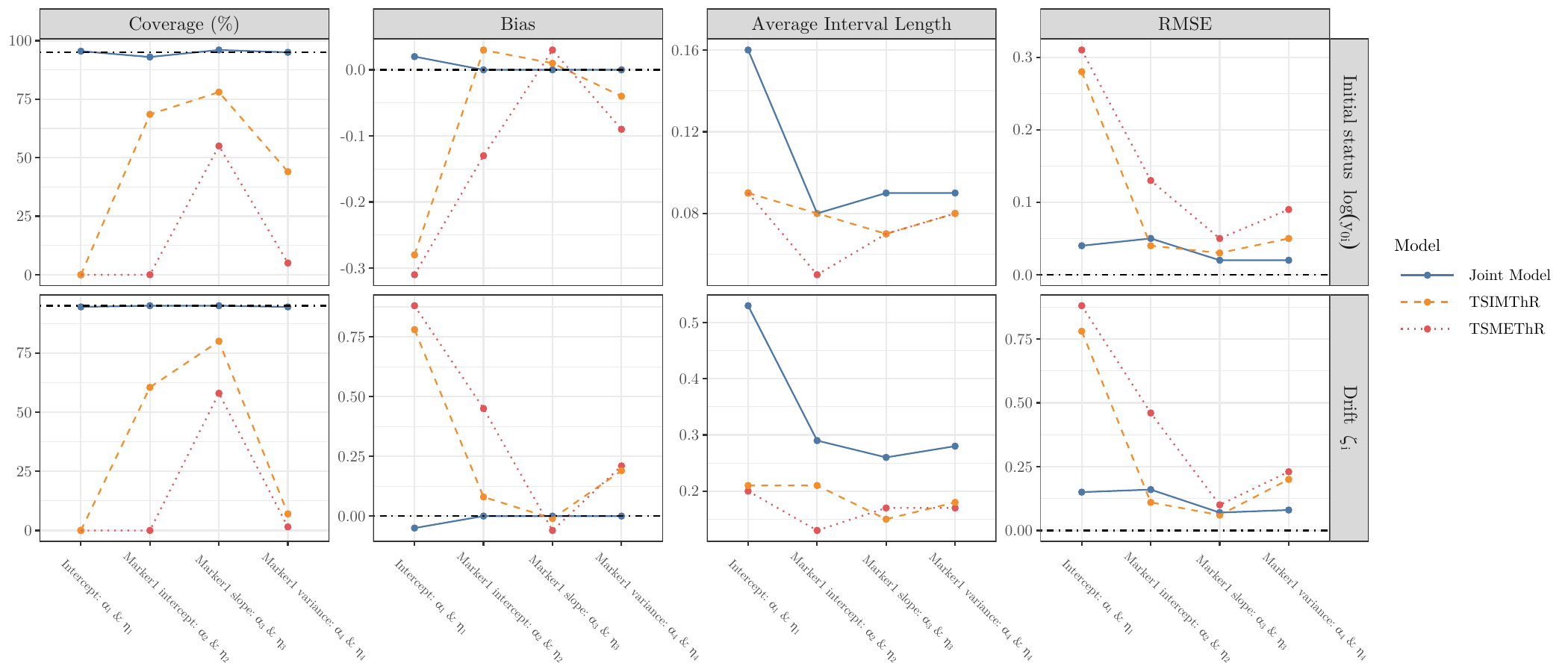}
\end{center}
\caption{\small Visualization of the simulation results with a single biomarker subject to a detection limit. The dot-dashed horizontal line in each panel indicates reference values: 95\% for Coverage (\%), 0 for Bias and RMSE. \label{fig:Q1_result}}
\end{figure}

Table \ref{tab:Q1_compare} and Figure \ref{fig:Q1_result} present the simulation results with a single biomarker subject to a detection limit. We again reach the same conclusion about the model comparison as of the main text and demonstrate the superiority of our proposed joint model.
Table \ref{tab:Q1_sim_supplement} and Table \ref{tab:Q2_sim_supplement} shows the simulation results of parameters in the longitudinal submodel (Component 1) estimated by the proposed joint model for both $Q=1$ with a detection limit and $Q=2$ with one biomarker subject to a detection limit. The results indicate that the joint model also achieves high coverage rates and very low bias when estimating parameters in the longitudinal biomarkers.

\begin{table}
\caption{\small Simulation results including coverage rate, bias, average interval length and RMSE of $\boldsymbol{\alpha}$ (regression coefficients of $\log(y_{0i})$) and $\boldsymbol{\eta}$ (regression coefficients of $\zeta_i$) and comparing the joint model with alternative two-stage approaches with a single biomarker subject to a detection limit.  \label{tab:Q1_compare}}
\renewcommand{\arraystretch}{1}
\begin{center}
\begin{tabular}{lrrrrr}
\hline
True Values & Model & Coverage (\%) & Bias & Average Interval Length & RMSE\\\hline
$\alpha_1$ = 3.5 & \textbf{Joint Model} & 95.5 & 0.02 & 0.16 & 0.04 \\ 
      & TSMEThR & 0.0 & -0.31 & 0.09 & 0.31 \\
      & TSIMThR & 0.0 & -0.28 & 0.09 & 0.28 \\
      $\alpha_2$ = 0.3 & \textbf{Joint Model} & 93.0 & -0.00 & 0.08 & 0.05 \\
      & TSMEThR & 0.0 & -0.13 & 0.05 & 0.13 \\
      & TSIMThR & 68.5 & 0.03 & 0.08 & 0.04 \\
    $\alpha_3$ = 0.2 & \textbf{Joint Model} & 96.0 & 0.00 & 0.09 & 0.02 \\
      & TSMEThR & 55.0 & 0.03 & 0.07 & 0.05 \\
      & TSIMThR & 78.0 & 0.01 & 0.07 & 0.03 \\
      $\alpha_4$ = 0.15 & \textbf{Joint Model} & 95.0 & -0.00 & 0.09 & 0.02 \\
      & TSMEThR & 5.0 & -0.09 & 0.08 & 0.09 \\
      & TSIMThR & 44.0 & -0.04 & 0.08 & 0.05 \\
      $\eta_1$ = -3 & \textbf{Joint Model} & 94.5 & -0.05 & 0.53 & 0.15 \\
      & TSMEThR & 0.0 & 0.88 & 0.20 & 0.88 \\
      & TSIMThR & 0.0 & 0.78 & 0.21 & 0.78 \\
      $\eta_2$ = -0.8 & \textbf{Joint Model} & 95.0 & 0.00 & 0.29 & 0.16 \\
      & TSMEThR & 0.0 & 0.45 & 0.13 & 0.46 \\
      & TSIMThR & 60.5 & 0.08 & 0.21 & 0.11 \\
      $\eta_3$ = -0.02 & \textbf{Joint Model} & 95.0 & 0.00 & 0.26 & 0.07 \\
      & TSMEThR & 58.0 & -0.06 & 0.17 & 0.10 \\
      & TSIMThR & 80.0 & -0.01 & 0.15 & 0.06 \\
      $\eta_4$ = -0.3 & \textbf{Joint Model} & 94.5 & -0.00 & 0.28 & 0.08 \\
      & TSMEThR & 1.5 & 0.21 & 0.17 & 0.23 \\
      & TSIMThR & 7.0 & 0.19 & 0.18 & 0.20 \\
      \hline
\end{tabular}
\end{center}
\end{table}

\begin{table}
\caption{\small Simulation results including coverage rate, bias, average interval length and RMSE of parameters in Component 1 when $Q=1$ with a detection limit.  \label{tab:Q1_sim_supplement}}
\renewcommand{\arraystretch}{1}
\begin{center}
\begin{tabular}{lrrrr}
\hline
True Values & Coverage (\%) & Bias & Average Interval Length & RMSE\\\hline
$\beta_{11}=6.5$ & 96.5 & -0.00 & 0.23 & 0.06 \\ 
    $\beta_{12} = 0.07$ & 96.5 & 0.00 & 0.07 & 0.02 \\
    $\beta_{13} = -0.06$ & 96.5 & -0.00 & 0.00 & 0.00 \\
    $\sigma_{11} = 0.75$ & 92.0 & -0.00 & 0.17 & 0.05 \\
    $\sigma_{12} = 0.3$ & 96.0 & 0.00 & 0.03 & 0.01 \\
    $\boldsymbol{\Omega}_1[1,2]=-0.1$ & 94.5 & 0.00 & 0.23 & 0.06 \\
    $\gamma_1=0.45$ & 93.5 & 0.00 & 0.15 & 0.04 \\
    $\Psi_1=1$ & 95.5 & 0.00 & 0.12 & 0.03 \\
      \hline
\end{tabular}
\end{center}
\end{table}

\begin{table}
\caption{\small Simulation results including coverage rate, bias, average interval length and RMSE of parameters in Component 1 when $Q=2$ with one biomarker subject to a detection limit.  \label{tab:Q2_sim_supplement}}
\renewcommand{\arraystretch}{1}
\begin{center}
\begin{tabular}{lrrrr}
\hline
True Values & Coverage (\%) & Bias & Average Interval Length & RMSE\\\hline
$\beta_{11} = 3$ & 93.5 & 0.00 & 0.16 & 0.04 \\ 
    $\beta_{12} = -0.2$ & 96.0 & 0.00 & 0.08 & 0.02 \\
    $\beta_{13} = 0.04$ & 95.5 & 0.00 & 0.01 & 0.00 \\
    $\beta_{14} = -0.001$ & 94.5 & 0.00 & 0.00 & 0.00 \\
    $\sigma_{11} = 0.23$ & 95.5 & 0.00 & 0.08 & 0.02 \\
    $\sigma_{12} = 0.05$ & 97.5 & 0.00 & 0.01 & 0.00 \\
    $\boldsymbol{\Omega}_1[1,2]=0.3$ & 96.5 & -0.01 & 0.43 & 0.10 \\
    $\gamma_1=-0.95$ & 99.5 & 0.00 & 0.08 & 0.02 \\
    $\Psi_1=0.45$ & 98.0 & -0.00 & 0.09 & 0.02 \\
    $\beta_{21} = 6.6$ & 96.5 & 0.00 & 0.22 & 0.05 \\ 
    $\beta_{22} = 0.03$ & 93.5 & -0.00 & 0.07 & 0.02 \\
    $\beta_{23} = -0.05$ & 93.0 & 0.00 & 0.00 & 0.00 \\
    $\sigma_{21} = 0.65$ & 96.0 & 0.00 & 0.16 & 0.04 \\
    $\sigma_{22} = 0.25$ & 98.0 & 0.00 & 0.03 & 0.01 \\
    $\boldsymbol{\Omega}_2[1,2]=0.18$ & 96.5 & 0.00 & 0.27 & 0.06 \\
    $\gamma_2=0.50$ & 93.5 & 0.01 & 0.15 & 0.04 \\
    $\Psi_2=1$ & 93.0 & -0.00 & 0.12 & 0.03 \\
    $a=5.3$ & 94.5 & 0.21 & 2.92 & 0.80 \\
    $b=12$ & 93.0 & 0.54 & 6.91 & 1.88 \\
      \hline
\end{tabular}
\end{center}
\end{table}

\section{Application to the Study of Women's Health Across the Nation (SWAN)}

In this section, we present the posterior estimates of the joint model from the case study, which is presented in Section 5 of the main text. Table \ref{tab:application_result} contains the estimates and 95\% credible intervals for parameters in the survival component, including model coefficients corresponding to the latent initial status $\log(y_{0i})$ and drift $\zeta_i$. Table \ref{tab:application_result_other} contains the estimates and 95\% credible intervals for parameters in the longitudinal component.

\begin{table}
\caption{\small Posterior mean estimates and 95\% credible intervals (CrI) for parameters in the survival component of the application model. The top panel corresponds to model results for the latent initial status $\log(y_{0i})$, and the bottom 
    panel corresponds to model results for the latent drift $\zeta_i$.  \label{tab:application_result}}
\renewcommand{\arraystretch}{1}
\begin{center}
\begin{tabular}{p{4cm}rrr}
\hline
Latent Component & Coefficient & Variable & Posterior Mean (95\% CrI) \\
       \hline
        \multirow{7}{*}{Initial status $\log(y_{0i})$}& $\alpha_1$ & Intercept & 3.54 (3.29, 3.81) \\
       & $\boldsymbol{\alpha_2}$ & \textbf{FSH intercept} & \textbf{-0.23 (-0.36, -0.14)} \\
       & $\boldsymbol{\alpha_3}$ & \textbf{FSH slope} & \textbf{0.19 (0.08, 0.36)} \\
      & $\alpha_4$ & FSH variance & -0.03 (-0.11, 0.05) \\
       & $\alpha_5$ & AMH intercept & 0.09 (-0.07, 0.22) \\
       & $\boldsymbol{\alpha_6}$ & \textbf{AMH slope} & \textbf{0.27 (0.16, 0.39)} \\
       & $\boldsymbol{\alpha_7}$ & \textbf{AMH variance} & \textbf{0.09 (0.02, 0.16)} \\
       & $\alpha_8$ & BMI & 0.00 (-0.00, 0.01) \\
       \hline
       \multirow{7}{*}{Drift $\zeta_i$} & $\eta_1$ & Intercept & -3.59 (-4.65, -2.72) \\
       & $\boldsymbol{\eta_2}$ & \textbf{FSH intercept} & \textbf{0.87 (0.48, 1.47)} \\
       & $\boldsymbol{\eta_3}$ & \textbf{FSH slope} & \textbf{-0.96 (-1.67, -0.53)} \\
        & $\eta_4$ & FSH variance & -0.03 (-0.33, 0.28) \\
       & $\eta_5$ & AMH intercept & -0.20 (-0.70, 0.38) \\
       & $\eta_6$ & AMH slope & -0.40 (-0.83, 0.00) \\
       & $\eta_7$ & AMH variance & -0.19 (-0.46, 0.06) \\
       & $\eta_8$ & BMI & -0.00 (-0.03, 0.02) \\
      \hline
\end{tabular}
\end{center}
\end{table}

\begin{table}
\caption{\small Posterior mean estimates and 95\% credible intervals (CrI) for parameters in the longitudinal component of the application model. \label{tab:application_result_other}}
\renewcommand{\arraystretch}{1}
\begin{center}
\begin{tabular}{lrr}
\hline
Parameter & Posterior Mean & Posterior 95\% CrI \\
       \hline
     $\beta_{11}$ & 2.96 &  (2.83, 3.10) \\
       $\beta_{12}$ & -0.20 & (-0.25, -0.14) \\
       $\beta_{13}$ & 0.035 & (0.030, 0.041) \\
      $\beta_{14}$ & -0.0009 & (-0.0011, -0.0007) \\
       $\sigma_{11}$ & 0.23 & (0.17, 0.30) \\
       $\sigma_{12}$ & 0.05 & (0.04, 0.06) \\
       $\boldsymbol{\Omega}_1[1,2]$ & 0.37 & (0.00, 0.79) \\
       $\gamma_1$ & -0.93 & (-0.98, -0.88) \\
       $\Psi_1$ & 0.46 & (0.40, 0.52) \\
       $\beta_{21}$ & 6.62 &  (6.43, 6.80) \\
       $\beta_{22}$ & 0.03 & (-0.02, 0.08) \\
       $\beta_{23}$ & -0.056 & (-0.059, -0.052) \\
       $\sigma_{21}$ & 0.64 & (0.46, 0.82) \\
       $\sigma_{22}$ & 0.25 & (0.23, 0.27) \\
       $\boldsymbol{\Omega}_2[1,2]$ & 0.18 & (-0.06, 0.48) \\
       $\gamma_2$ & 0.49 & (0.40, 0.57) \\
       $\Psi_2$ & 0.99 & (0.92, 1.07) \\
       $a$ & 5.34 & (3.80, 7.67) \\
       $b$ & 12.27 & (8.62, 17.74) \\
      \hline
\end{tabular}
\end{center}
\end{table}

\section{Posterior Predictive Model Checking}

To diagnose the imputation procedure and assess the validity of our model construction on the SWAN data, we conduct a posterior predictive check (PPC) on both the longitudinal biomarker component and the time-to-event outcome component. In the case of fully observed data without missing values, let $\boldsymbol{Y}$ denote the observed data and $\boldsymbol{\Theta}$ the unknown model parameters. The replicated data $\boldsymbol{Y}^{\text{rep}}$ can be simulated from its posterior predictive distribution $\mathcal{P}(\boldsymbol{Y}^{\text{rep}}|\boldsymbol{Y}, \boldsymbol{\Theta})$. Comparisons between observed and replicated data are made using test statistics $T(\boldsymbol{y}, \boldsymbol{\Theta})$. However, since missing values and imputations are present in both the longitudinal and survival submodels, the PPC must be adapted to incorporate diagnostics for the imputation models. Let $\boldsymbol{Y}_{\text{com}}$ represent the completed data, which consists of the observed data $\boldsymbol{Y}_{\text{obs}}$ and missing data $\boldsymbol{Y}_{\text{mis}}$, and let $\boldsymbol{Y}^{\text{rep}}_{\text{com}}$ represent the replication of the completed data. The imputation of $\boldsymbol{Y}_{\text{mis}}$ and the replicated data $\boldsymbol{Y}^{\text{rep}}_{\text{com}}$ can be drawn from their posterior predictive distribution, $\mathcal{P}(\boldsymbol{Y}^{\text{rep}}_{\text{com}}, \boldsymbol{Y}_{\text{mis}}|\boldsymbol{Y}_{\text{obs}})=\int \mathcal{P}(\boldsymbol{Y}^{\text{rep}}_{\text{com}}|\boldsymbol{\Theta})\mathcal{P}(\boldsymbol{Y}_{\text{mis}}|\boldsymbol{\Theta}, \boldsymbol{Y}_{\text{obs}})\mathcal{P}(\boldsymbol{\Theta}|\boldsymbol{Y}_{\text{obs}})d\boldsymbol{\Theta}$. Test statistics are then calculated accordingly to compare the completed data with the replicated data.

\subsection{Posterior Predictive Check for Longitudinal Submodel}

The test statistic used to evaluate the longitudinal submodel is a chi-squared discrepancy statistic, defined as $T(\boldsymbol{X}_{i\cdot q};\boldsymbol{\Theta})=\sum_{j=1}^{n_i}(X_{ijq}-\mu_{ijq})^2/s_{iq}^2, q=1,2$, where $X_{ijq}$ represents the value of the $q$-th biomarker for the $i$-th individual at time $t_{ij}$ with $\mu_{ijq}$ being the corresponding mean function and $s_{iq}^2$ being the residual variance. The imputed values and replicated data are simulated from the following procedure:
\begin{enumerate}
    \item Simulate $K$ draws of $\boldsymbol{\Theta}$ from its posterior distribution.
    \item For each $\boldsymbol{\Theta}^k, k=1,\ldots,K$, the imputed value $X_{\text{mis},ijq}^k$, $q=1,2$, more specifically $\text{AMH}_{\text{mis},ij}^k$, is simulated from $\text{AMH}_{\text{mis},ij}^k|\text{FSH}_{\text{obs},ij}=x_{ij1}\sim \mathcal{N}(\mu_{ij2}^k+r_i^ks_{i2}^k(x_{ij1}-\mu_{ij1}^k)/s_{i1}^k, (1-(r_i^k)^2)(s_{i2}^k)^2;\text{AMH}_{\text{mis},ij}^k<\text{LOD}), j=1,\ldots,n_i,i=1,\ldots,N$. Here, $\mu_{ij1}^k$ and $\mu_{ij2}^k$ represent the mean function of $\text{FSH}_{\text{obs},ij}$ and $\text{AMH}_{\text{mis},ij}^k$, respectively. Similarly, $s_{i1}^k$ and $s_{i2}^k$ represent the residual standard deviation of the corresponding biomarker and $r_i^k$ represents the correlation between the two biomarkers. The above-mentioned parameters are all based on the $k$-th draw of $\boldsymbol{\Theta}$.
    \item For each $\boldsymbol{\Theta}^k, k=1,\ldots,K$, the replicated data $\boldsymbol{X}_{\text{com},ij}^{\text{rep},k}$ are simulated from $\mathcal{P}(\boldsymbol{X}_{\text{com},ij}^{\text{rep},k}|\boldsymbol{\Theta}^k)$ as $\boldsymbol{X}_{\text{com},ij}^{\text{rep},k}\sim \mathcal{N}_2(\boldsymbol{\mu}(t_{ij};\boldsymbol{\beta}_1^k, \boldsymbol{\beta}_2^k, \boldsymbol{B}_i^k), \boldsymbol{S}_i^k), j=1,\ldots,n_i, i=1,\ldots,N$.
\end{enumerate}
We simulate $K=2,000$ imputed values and replicated data for each individual. For each biomarker $q=1,2$ and individual $i=1,\ldots,N$, we compute 2,000 values of the test statistic $T(\boldsymbol{X}_{\text{obs},i\cdot q},\boldsymbol{X}_{\text{mis},i\cdot q}^k; \boldsymbol{\Theta})$ and $T(\boldsymbol{X}_{\text{com},i\cdot q}^{\text{rep},k}; \boldsymbol{\Theta})$. These two sets of test statistics are compared using a posterior predictive p-value, which quantifies the probability that the replicated data are more extreme than the completed data. The posterior predictive p-value is defined as $\mathcal{P}(T(\boldsymbol{X}_{\text{obs},i\cdot q},\boldsymbol{X}_{\text{mis},i\cdot q}^k; \boldsymbol{\Theta})\le T(\boldsymbol{X}_{\text{com},i\cdot q}^{\text{rep},k}; \boldsymbol{\Theta})|\boldsymbol{X}_{\text{obs},i\cdot q})$. An extreme p-value close to 0 or 1 implies poor model fit \citep{gelman2013bayesian}.

\begin{figure}
\begin{center}
\includegraphics[width=5in]{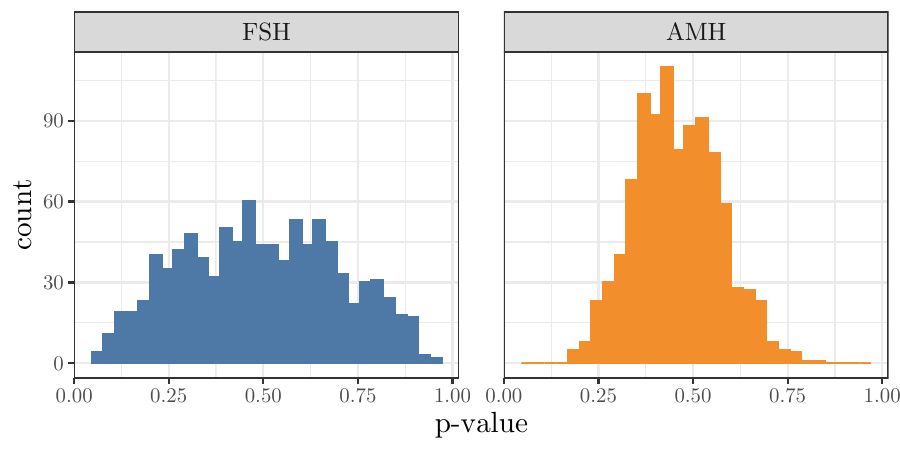}
\end{center}
\caption{\small Posterior predictive p-values of checking FSH and AMH trajectories across all subjects for the longitudinal submodel. \label{fig:PPC_long}}
\end{figure}

Figure \ref{fig:PPC_long} displays histograms of the posterior predictive p-values for each individual's FSH and AMH trajectories. The median of the posterior predictive p-values for checking FSH and AMH trajectories are 0.49 and 0.45, respectively. 72.4\% of the p-values for FSH and 96.7\% for AMH fall within the range of 0.25 to 0.75. These results indicate that our imputation procedure generates reasonable imputed values and that the longitudinal submodel adequately fits the observed hormone trajectories.

\subsection{Posterior Predictive Check for Survival Submodel}
\begin{figure}
\begin{center}
\includegraphics[width=5in]{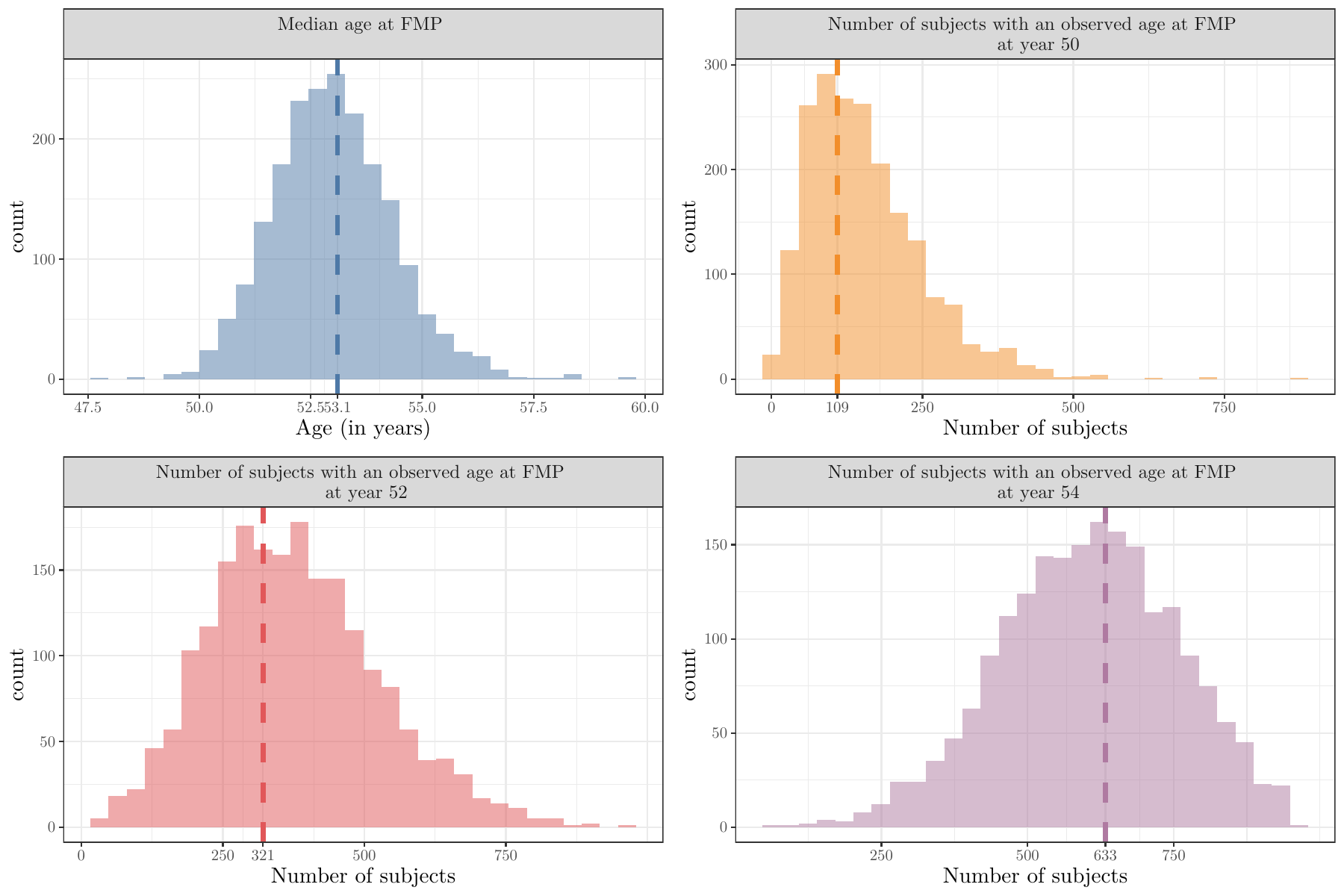}
\end{center}
\caption{\small Posterior predictive check for the survival component. The top left panel shows the histogram of the median age at FMP from the replicated data. The dashed line represents the median age at FMP from the completed data when taking the posterior means as the imputed event times. The top right, bottom left, and bottom right panels correspond to the histogram of the number of individuals with an observed FMP at years 50, 52, and 54, respectively, from the replicated data. The dashed lines represent the same measure from the completed data when taking the posterior means as the imputed event times. \label{fig:PPC_surv}}
\end{figure}

A similar procedure is used to evaluate the survival submodel, as we impute the event times for subjects whose age at FMP is right-censored. The test statistics we employ are the median age at FMP and the number of subjects with an observed age at FMP at 50, 52, and 54 years old. The imputed age at FMP and the replicated ones are simulated from the following procedure:
\begin{enumerate}
    \item Simulate $K$ draws of $\boldsymbol{\Theta}$ from its posterior distribution.
    \item For each $\boldsymbol{\Theta}^k, k=1,\ldots,K$, the imputed event time $T_{\text{mis},i}^k$ is simulated from $T_{\text{mis},i}^k|C_i=c_i\sim \text{Inverse-Gaussian}(-y_{0i}^k/\zeta_i^k,(y_{0i}^k)^2; T_{\text{mis},i}^k>c_i)$ when $\delta_i=0$.
    \item For each $\boldsymbol{\Theta}^k, k=1,\ldots,K$, the replicated data are simulated as $T_{\text{com},i}^{\text{rep},k}\sim \text{Inverse-Gaussian}($\\$-y_{0i}^k/\zeta_i^k$, $(y_{0i}^k)^2), i=1,\ldots,N$.
\end{enumerate}
Again, we simulate $K=2,000$ imputed values and replicated data. The test statistics are calculated for each completed and replicated dataset and compared using the posterior predictive p-values. Figure \ref{fig:PPC_surv} displays the histograms of the median age at FMP and the number of subjects with an observed FMP at ages 50, 52, and 54 for the replicated data. The histograms are centered around the corresponding measures from the completed data when we take the posterior means as the imputed event times. The posterior predictive p-value for comparing the median age at FMP between the replicated data and the completed data is 0.47, and the posterior predictive p-values for comparing the number of subjects with an observed FMP at ages 50, 52, and 54 are 0.63, 0.61 and 0.43, respectively. These results again demonstrate the validity of the model structure and imputation procedure for the time-to-event outcome.

\bibliographystyle{apalike}

\bibliography{references}